\documentclass[a4paper,fleqn,usenatbib]{mnras}

\usepackage{newtxtext,newtxmath}

\usepackage[T1]{fontenc}
\usepackage{ae,aecompl}

\usepackage{graphicx}	
\usepackage{amssymb}	

\usepackage{tabularx, dcolumn }

\newcolumntype{d}[1]{D{.}{.}{#1}}

\title[Radio jets associated with quasar feedback]{Prevalence of radio jets associated with galactic outflows and feedback from quasars}

\author[M.\,E.\,Jarvis et al.]{M.\,E.\,Jarvis,$^{1,2,3}$\thanks{E-mail: miranda.jarvis@gmail.com}
C.\,M.\.Harrison,$^{2}$
A.\,P.\,Thomson,$^{4}$
C.\,Circosta,$^{2,3}$
V.\,Mainieri,$^{2}$ \newauthor
 D.\,M.\,Alexander,$^{5}$
 A.\,C.\,Edge,$^{5}$
 G.\,B.\,Lansbury,$^{6}$
 S.\,J.\,Molyneux$^{2}$
and J.\,R.\,Mullaney$^{7}$
\\
$^{1}$Max-Planck Institut f\"ur Astrophysik, Karl-Schwarzschild-Str. 1, 85748 Garching, Germany \\
$^{2}$European Southern Observatory, Karl-Schwarzschild-Str. 2, 85748 Garching, Germany \\
$^{3}$Ludwig Maximilian Universit\"at, Professor-Huber-Platz 2, 80539 Munich, Germany\\
$^{4}$Jodrell Bank Centre for Astrophysics, The University of Manchester, Oxford Road, Manchester M13 9PL, UK \\
$^{5}$Centre for Extragalactic Astronomy, Department of Physics, Durham University, South Road, Durham DH1 3LE, UK\\
$^{6}$Institute of Astronomy, University of Cambridge, Madingley Road, Cambridge, CB3 0HA, UK\\
$^{7}$Department of Physics and Astronomy, The University of Sheffield, Hounsfield Road, Sheffield, S3 7RH, UK
}

\date{Accepted XXX. Received YYY; in original form ZZZ}

\pubyear{2018}

\begin{document}
\label{firstpage}
\pagerange{\pageref{firstpage}--\pageref{lastpage}}
\maketitle



\begin{abstract}

We present 1--7~GHz high-resolution radio imaging (VLA and e-MERLIN) and spatially-resolved ionized gas kinematics for ten $z<0.2$ type~2 `obscured' quasars ($\log [L_{\text{AGN}}$/erg\,s$^{-1}]$$\gtrsim$$45$) with moderate radio luminosities ($\log [L_{\text{1.4\,GHz}}$/W\,Hz$^{-1}]$=23.3--24.4). These targets were selected to have known ionized outflows based on broad [O~{\sc iii}] emission-line components (FWHM$\approx$800--1800\,km\,s$^{-1}$). Although `radio-quiet' and not `radio AGN' by many traditional criteria, we show that for nine of the targets, star formation likely accounts for $\lesssim$10~per\,cent of the radio emission. We find that $\sim$80--90~per cent of these nine targets exhibit extended radio structures on 1--25~kpc scales. The quasars' radio morphologies, spectral indices and position on the radio size-luminosity relationship reveals that these sources are consistent with being low power compact radio galaxies. Therefore, we favour radio jets as dominating the radio emission in the majority of these quasars. The radio jets we observe are associated with morphologically and kinematically distinct features in the ionized gas, such as increased turbulence and outflowing bubbles, revealing jet-gas interaction on galactic scales. Importantly, such conclusions could not have been drawn from current low-resolution radio surveys such as FIRST. Our observations support a scenario where compact radio jets, with modest radio luminosities, are a crucial feedback mechanism for massive galaxies during a quasar phase. 


\end{abstract}




\section{Introduction}



Growing supermassive black holes at the hearts of massive galaxies, i.e., active galactic nuclei (AGN), are widely believed to be able to impact galaxy evolution by facilitating a global shut-down or regulation of star formation \citep[e.g., see reviews in] []{Alexander12,Fabian12,Harrison17}. Galaxy formation models require this `AGN feedback' to inject energy or momentum into the surrounding gas, in order to reproduce key observables of galaxy populations and the intergalactic material \citep[e.g.,][]{2006ApJS..163....1H, 2006MNRAS.370..645B,2010MNRAS.406..822M,Gaspari11,Dubois13,Vogelsberger14,Hirschmann14,2015MNRAS.446..521S,Henriques15,TaylorKobayashi15,Choi18}. 

Historically, `AGN feedback' was considered to come in two flavours: `quasar mode', and `maintenance mode' or `radio mode' \cite[e.g., see] []{Croton09,2012MNRAS.422.2816B}. The former mode is associated with powerful radiatively-dominated AGN, which are often referred to as quasars \cite[e.g., see] []{Harrison17}. The energetic photons are predicted to couple to the nearby gas, resulting in high-velocity winds that propagate through the host galaxy and ultimately in the removal or destruction of the star forming fuel \citep[e.g.,][]{FaucherGiguere12,King15,Costa18}. Conversely, maintenance mode is associated with low accretion rate AGN that release most of their energy in the form of radio jets \citep{McNamara12}. These jets regulate the cooling of gas in the halos, and hence the level of star formation in their host galaxies. In reality feedback is unlikely to be simply divided into two modes \citep[e.g.,][]{2005MNRAS.363L..91C, 2010ApJ...717..708C,Cielo+18} and observations are ultimately required to determine the processes by which AGN impact upon their galaxies. 

Observationally, the details of `quasar mode' feedback are not clear. On the one hand, winds driven in the vicinity of the accretion disc are common, if not ubiquitous \citep[e.g.,] []{SilkRees98,King15} and multi-phase AGN-driven outflows have been observed on galaxy-wide scales (i.e., $\gtrsim$0.5\,kpc) using integral field spectroscopy (IFS) and interferometric observations \cite[e.g., ][]{Veilleux13,Husemann13,Liu13,2013ApJ...768...75R,Cicone14,Harrison14,Fiore17,Bae18,Morganti18,Fluetsch18}. However, determining how these galactic-outflows are driven is challenging; with accretion-disc winds, radio jets and star-formation all potential candidates \cite[e.g.,][]{Harrison18,Wylezalek18}.

Whilst radio jets are unambiguously associated with galaxy-wide outflows in rare, extremely radio luminous quasars \citep[i.e., $L_{\rm 1.4GHz}$$>$10$^{25}$~\,W\,Hz$^{-1}$;][]{Nesvadba17}; the majority of quasars ($\gtrsim$90~per cent) have lower radio luminosities. Particularly at moderate luminosities (i.e., 10$^{23}$$\lesssim$$L_{\rm 1.4GHz}$$\lesssim$$10^{25}$~\,W\,Hz$^{-1}$), the dominant origin of radio emission is a matter of ongoing debate \citep[e.g.,][]{Condon13,Padovani15,2004AJ....128.1002Z,Zakamska16}. Furthermore, studies using spatially-unresolved radio emission and spectroscopy are unable to definitively distinguish between winds and radio jets as driving galactic outflows in typical quasars \citep{Mullaney13,VillarMartin14,Zakamska14}.



As part of an ongoing programme, in this work we use spatially-resolved radio observations and spectroscopy to assess the dominant producer of radio emission and drivers of galactic ionized outflows in quasars. Using a sample of $\sim$24,000 $z<0.4$ AGN we already discovered a strong relationship between the radio luminosity and the prevalence of ionized outflows based on measuring the [O~{\sc iii}] emission-line profiles \citep[\citealt{Mullaney13}; also see][]{Zakamska14,VillarMartin14}. Here, we combine follow-up high-resolution radio observations and integral field spectrograph observations of ten $z<0.2$ quasars with moderate radio luminosities.

In Section~\ref{selection&characterisation} we describe the sample selection criteria and characterise the sample's host galaxy and AGN properties, using spectral energy distributions (SEDs). In Section~\ref{obs&redux} we describe the radio and IFS data sets we used and reduction steps taken and in Section~\ref{analysis} we describe the details of our analyses. In Section~\ref{results} we discuss our results in the context of previous work. Finally, in Section~\ref{conclusions} we give our conclusions. We adopt $H_0$=71km s$^{-1}$Mpc$^{-1}$, $\Omega_M$=0.27, $\Omega_\Lambda$=0.73 throughout, and define the radio spectral index, $\alpha$, using $S_\nu \propto \nu^{\alpha}$. We assume a \citet{Chabrier03} initial mass function (IMF).


\section{Target selection and characterisation}
\label{selection&characterisation}

\subsection{Sample selection}
\label{selection}

\begin{figure}
 \centering
 \includegraphics[width=\hsize]{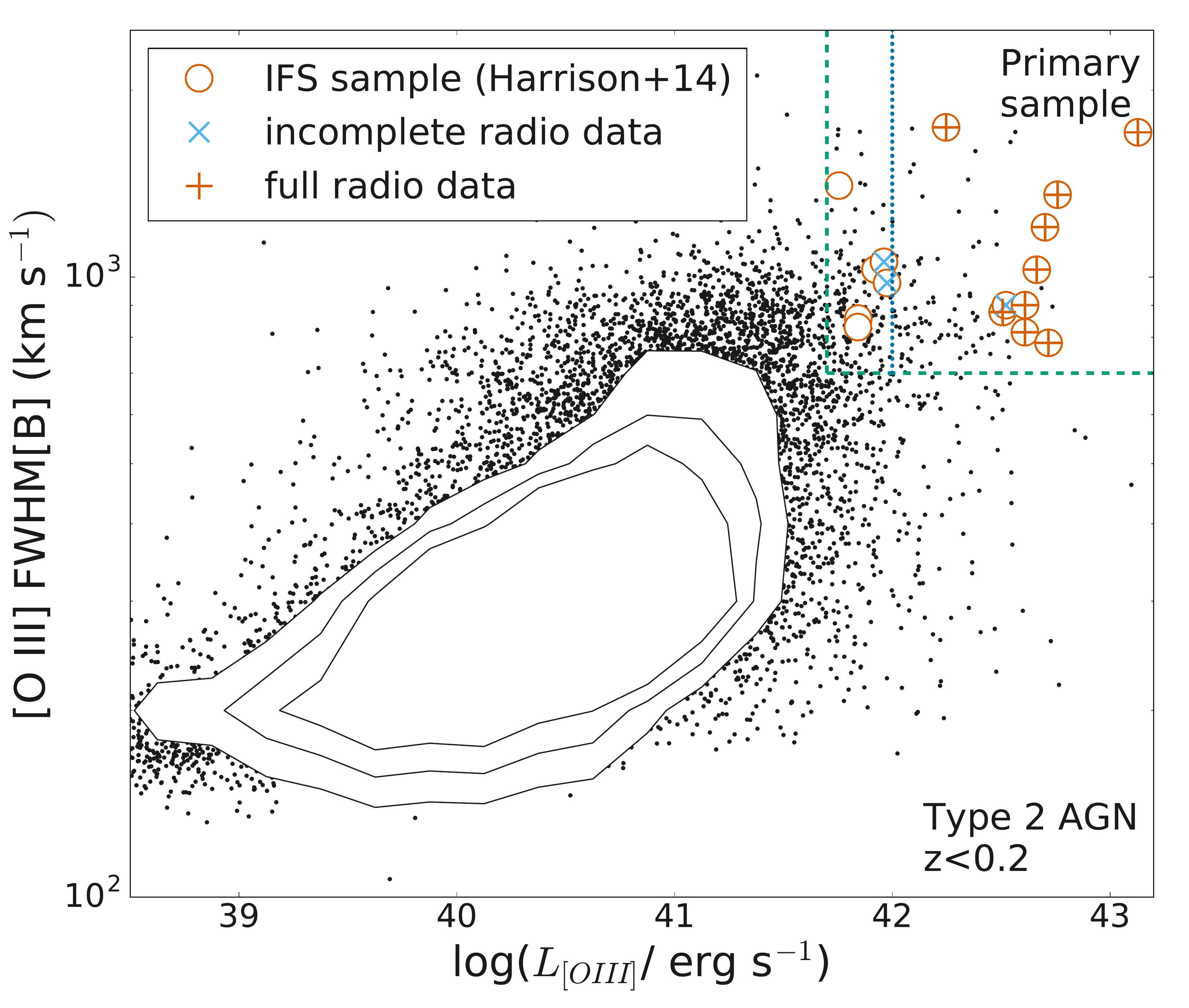}
 \caption{ The FWHM of the broadest, luminous [O~{\sc iii}] emission-line component versus the total [O~{\sc iii}] luminosity (see Table~\ref{targets} and Section~\ref{selection}). Our parent population of $z<0.2$ type 2 AGN are shown as black data points and contours \citep{Mullaney13}. The dashed green lines show the selection criteria used in \citet{Harrison14} to select sources with spectral signatures of ionized outflows, the dotted blue line marks the additional criteria for the primary sample considered for most of the analysis presented here ($L_{\rm [O~III]}>10^{42}$\,erg\,s$^{-1}$). The IFS targets are shown as red circles, with red plus symbols marking those with full radio data and cyan crosses marking those with incomplete radio data (see Section~\ref{obs&redux}).  }
  \label{sample}
 \end{figure}
\begin{figure*}
 \centering
 \includegraphics[width=18cm]{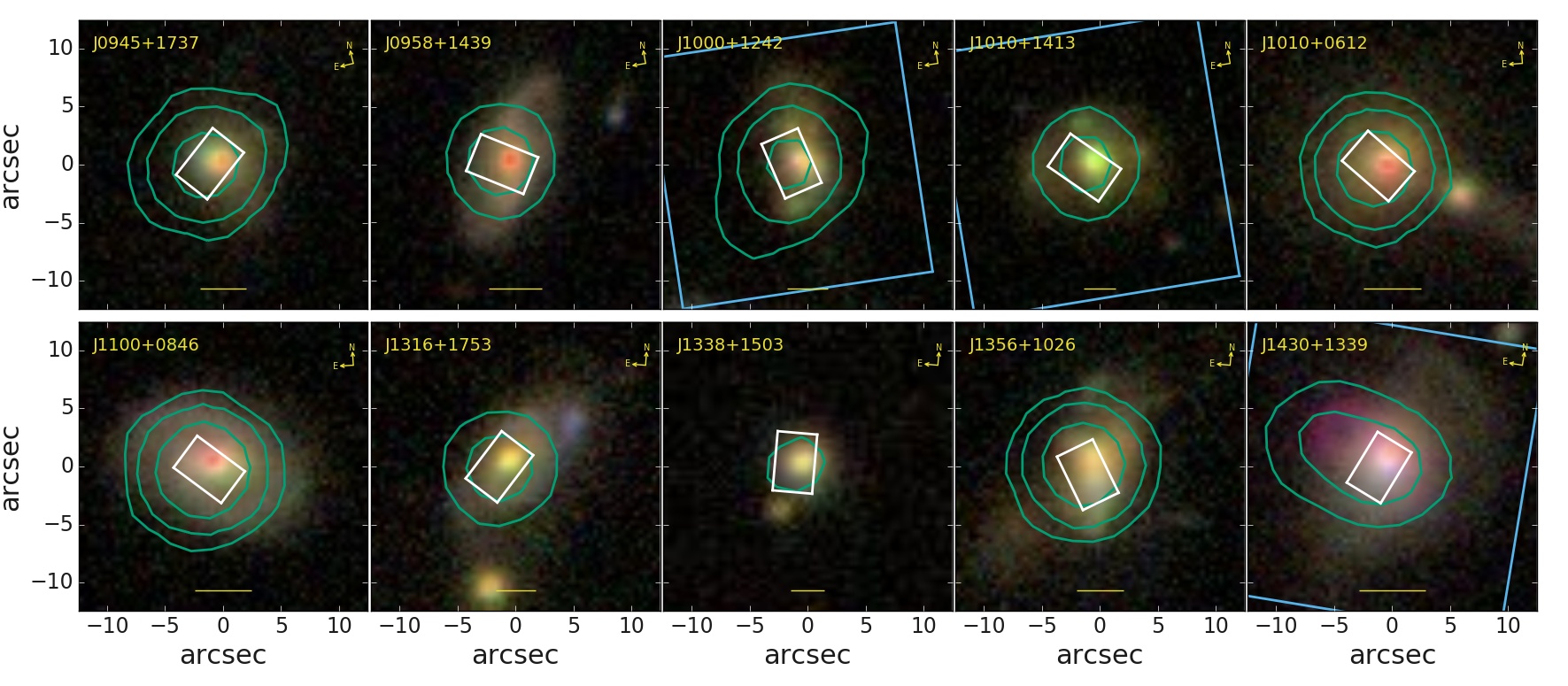}
 \caption{An overview of the ten quasars presented here. A three-colour (\emph{gri}) SDSS image is shown \citep{2015ApJS..219...12A}. The green contours show the FIRST 1.4\,GHz, $\sim$5~arcsec resolution data, with contours at $\pm[8,32,128]\sigma$. The rectangles mark the field of view of our IFS data (GMOS in white and VIMOS in blue). The scale bar in each marks 9\,kpc. Based on the FIRST data (also see Table~\ref{targets}), only one source shows unambiguous extended radio emission (J1430$+$1339) and two show marginal evidence for extended emission (J0945$+$1737; J1000$+$1242) and in all cases the resolution is too poor to connect to the ionized gas kinematics observed in the IFS data. }
  \label{sdss}
 \end{figure*}

In this work we focus on ten type~2 (`obscured') $z<0.2$ AGN, that have quasar-like luminosities \citep[i.e., $L_{\rm [O~III]}$$>$$10$$^{42}$\,erg\,s$^{-1}$;][]{Reyes+08}. These were originally selected by \citet{Harrison14} from our parent sample of 24\,264 $z<0.4$ spectroscopically identified AGN presented in \citet{Mullaney13}. We originally selected 16 sources for follow-up IFS observations that exhibit a luminous broad [O~{\sc iii}] component in the one dimensional spectra, indicative of a powerful ionized outflow \citep[see Fig.~\ref{sample};][]{Harrison14}. These IFS data revealed $\sim$kpc scale ionised outflows. The present text focuses on the subset of ten of these targets with a luminosity of $L_{\rm [O~III]}>10^{42}$\,erg\,s$^{-1}$ (referred to as the primary sample; see Fig.~\ref{sample}). In Fig.~\ref{sdss} we show three-colour SDSS images for the ten quasars discussed in this work with the IFS field of view over-plotted.

%

%

\begin{figure*}
 \centering
 \includegraphics[width=18cm]{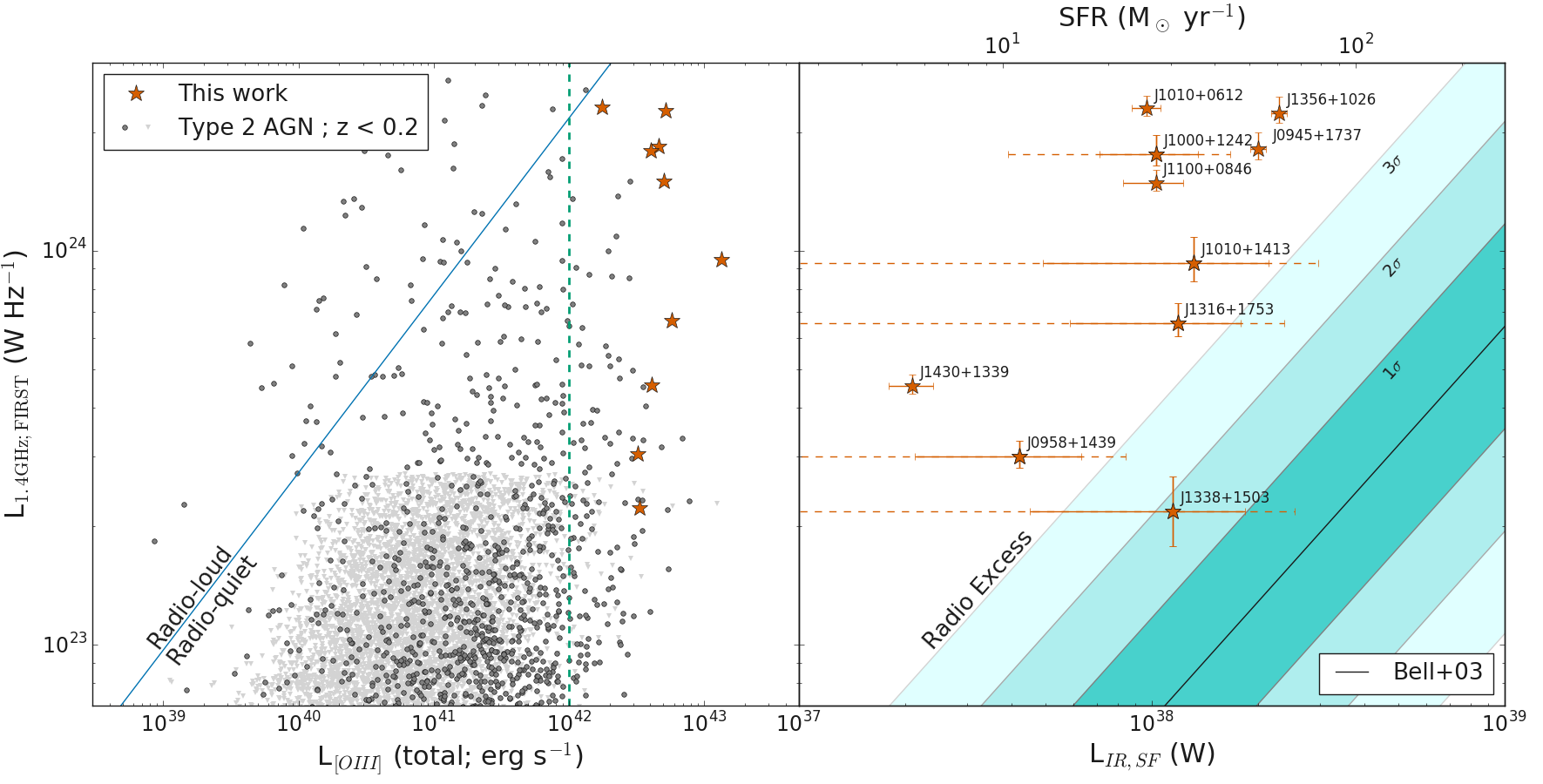}
 \caption{ Left: Radio luminosity (from FIRST fluxes) versus [O III] luminosity for our galaxies (red stars) with the division between `radio-loud' and `radio-quiet' AGN from \citet{Xu+99} (blue line; see Section~\ref{selection}). Also shown is our parent sample of type 2 AGN with $z<0.2$ \citep[black points, grey for upper limits;][]{Mullaney13}. Our selection criterion of $L_{\rm [O III]} > 10^{42}$ \,erg\,s$^{-1}$ is shown as a dashed green line. Right: The \emph{FIR}--radio correlation of \citet{Bell+03} compared to the values for our primary sample. We plot the far-infrared luminosity of the star formation component only ($L_{IR,SF}$). For those sources without good \emph{FIR} coverage the 2$\sigma$ errors are also plotted as dashed lines (see Section~\ref{characterising} and Table~\ref{SED_vals}). The solid black line is the average correlation from \citet{Bell+03} with the cyan regions marking 1, 2 and 3$\sigma$ regions respectively. Although all of our targets are classified as radio-quiet, all but one are consistent with having excess radio emission above that predicted from star formation alone (see Section~\ref{characterising}). 
  }
  \label{radio_loudness}
 \end{figure*}

The positions, redshifts, [O~III] properties and radio properties (from the FIRST Survey; \citealt{1995ApJ...450..559B}) of the ten targets studied in this paper are presented in Table~\ref{targets}. As can be seen in Fig.~\ref{radio_loudness} all of the primary sample discussed here are classified as `radio-quiet' AGN based upon the criteria of \citet{Xu+99}. 
Furthermore, as can be seen from the radio contours from FIRST in Fig.~\ref{sdss}, the spatial resolution ($\sim$5~arcsec) of these data, compared to our IFS observations (i.e., $\sim$0.6--0.9~arcsec; Section~\ref{IFS}), is insufficient to unambiguously relate the kinematic features observed in the IFS data to the radio morphology. This motivated us to observe all of our targets with interferometric radio observations (as described in Section~\ref{obs&redux}) to obtain higher resolution radio images \citep[also see][]{Harrison15}. 



\begin{table*}
 \caption{Target list and basic properties.
 \newline Notes: (1) Object name; (2)-(3) optical RA and DEC positions from SDSS (DR7); (4) Systemic redshifts from the GMOS data (see Section~\ref{line_fitting}); (5) Total observed [O~{\sc iii}]$\lambda5007$ luminosity from \citet{Mullaney13}. Absorption corrections would increase the values by on average 0.6\,dex (with a maximum increase of 1.4\,dex); (6) FWHM of the broad component of the [O~{\sc iii}] line fit from \citet{Mullaney13}; (7) 1.4~GHz flux densities obtained from the FIRST survey \citep{1995ApJ...450..559B} and uncertainties that
are defined as 3$\times$ the RMS noise of the radio image at the source position; (8) Rest-frame radio luminosities using a spectral index of $\alpha=-0.7$ and assuming $S_\nu \propto \nu^{\alpha}$ (we note that a range of $\alpha=-0.2$ to $-1.5$ introduces a spread of $\pm$ 0.1 dex on the radio luminosity); 
(9) Radio morphology parameter, where sources with $\Theta>1.06$ are classified as extended in the 1.4~GHz FIRST data \citep{Harrison14}. }

	\begin{tabular}{c c c c c d{4.0} d{2.2} c l} 
	\hline
   
	 Name & RA (J2000) & Dec (J2000) & z & $\log(L_{\text{[O~{\sc iii}]}})$ & \multicolumn{1}{c}{FWHM[B]} &  \multicolumn{1}{c}{$S_{1.4}$}& $\log(L_{1.4})$ & \multicolumn{1}{c}{$\Theta_{\text{FIRST}}$}\\ 
   & & & & (erg s$^{-1})$ & \multicolumn{1}{c}{(km s$^{-1}$)} & \multicolumn{1}{c}{(mJy) } & (W Hz$^{-1}$) & \\
 (1) & (2) & (3) & (4) & (5) & \multicolumn{1}{c}{(6)} & \multicolumn{1}{c}{(7) }& (8) & \multicolumn{1}{c}{(9)}\\
	\hline   
J0945$+$1737 & 09:45:21.33 & +17:37:53.2 & 0.1281 & 42.66 & 1027 & 44.5[4] & 24.3 & 1.072[3] \\
J0958$+$1439 & 09:58:16.88 & +14:39:23.7 & 0.1091 & 42.51 & 878 & 10.4[4] & 23.5 & 1.006[9] \\
J1000$+$1242 & 10:00:13.14 & +12:42:26.2 & 0.1479 & 42.61 & 815 & 31.8[4] & 24.2 & 1.111[4] \\
J1010$+$1413 & 10:10:22.95 & +14:13:00.9 & 0.1992 & 43.13 & 1711 & 8.8[5] & 24.0 & 1.04[1] \\
J1010$+$0612 & 10:10:43.36 & +06:12:01.4 & 0.0982 & 42.25 & 1743 & 99.3[3] & 24.4 & 1.038[3] \\
J1100$+$0846 & 11:00:12.38 & +08:46:16.3 & 0.1004 & 42.70 & 1203 & 61.3[3] & 24.2 & 1.023[1] \\
J1316$+$1753 & 13:16:42.90 & +17:53:32.5 & 0.1504 & 42.76 & 1357 & 11.4[4] & 23.8 & 1.04[1] \\
J1338$+$1503 & 13:38:06.53 & +15:03:56.1 & 0.1859 & 42.52 & 901 & 2.4[4] & 23.3 & 1.06[5] \\
J1356$+$1026 & 13:56:46.10 & +10:26:09.0 & 0.1233 & 42.72 & 783 & 59.6[4] & 24.4 & 1.014[2] \\
J1430$+$1339 & 14:30:29.88 & +13:39:12.0 & 0.0852 & 42.61 & 901 & 26.4[4] & 23.7 & 1.400[9] \\
\hline

	\end{tabular}
    
\label{targets}

	\end{table*}
%


\subsection{Star-formation rates and SED fitting}
\label{characterising}

In Fig.~\ref{radio_loudness} we show that all of the quasars in our sample would be classified as `radio-quiet' by the \citet{Xu+99} criterion. However, there is significant debate in the literature as to whether this division marks two populations or one continuous distribution and if these divisions are physically motivated \citep{Cirasuolo03,2017NatAs...1E.194P,Padovani17,Gurkan18}. A more meaningful measure is to select `radio AGN' by assessing if the observed radio emission is dominated by star formation or by the AGN \citep[e.g.][]{Moric10,Best&Heckman}. Although our sources would be classified as star forming by the method of \citet{Best&Heckman}, the method that we use in this work is to look for an excess of radio emission in relation to the \emph{FIR}--radio correlation for star-forming galaxies \citep[e.g.][]{Helou+85,Bell+03}.

An important consideration in using the \emph{FIR}--radio correlation to identify so called `radio excess' galaxies is separating the \emph{FIR} contribution from star formation and the AGN. This is because if both the \emph{FIR} emission and radio emission are dominated by the AGN, this could produce another correlation, artificially causing AGN to follow the relation set by star-forming galaxies \citep{Moric10,Zakamska16}. We therefore make use spectral energy distribution (SED) fitting from the \emph{UV} to \emph{FIR} to isolate the \emph{FIR} luminosity associated with star formation ($L_{\rm IR,SF}$) in addition to getting stellar masses ($M_{\star}$) and AGN bolometric luminosities ($L_{\rm AGN}$). The details of the archival photometric data we used are provided in the online supplementary information (Appendix A). We note that five of our targets do not have photometric measurements at wavelengths $\gtrsim$60$\mu$m; we flag these targets in Table~\ref{SED_vals} and assess the reliability of our key parameters for these targets below.

To fit the SEDs we used the Code Investigating GALaxy Emission \citep[CIGALE\footnote{\url{https://cigale.lam.fr}};][]{noll09,buat15,ciesla15}. We followed the basic procedure described in \citet{Circosta18} but provide specific details of our implementation of the code in the online supplementary information (Appendix A). In short, the code simultaneously fits attenuated stellar emission, dust emission heated by star formation, AGN emission (both primary accretion disc emission and dust heated emission) and nebular emission from the \emph{UV} to \emph{FIR}. The code builds up a probability distribution function (PDF) for each parameter of interest, taking into account the variations from the different models. These fits are an improvement on those previously presented in \citet{Harrison14}, most notably, the increased wavelength range used allowed the attenuated stellar emission and dust emission due to star formation to be coupled, increasing the accuracy of $L_{\rm IR,SF}$, particularly in the cases with limited \emph{FIR} coverage. Our results show that our targets have $L_{\rm IR,SF}$ that are broadly consistent with luminous infrared galaxies (i.e., 10$^{11}$\,$L_{\odot}$$\lesssim$$L_{\rm IR,SF}$$\lesssim$$10^{12}$\,$L_{\odot}$).

SED fits, especially with limited FUV coverage and an AGN component, can be difficult to determine \citep[see e.g.,][]{Bongiorno12,Zakamska16}. Indeed, it is well known that the uncertainties from standard SED-fitting procedures can be artificially small, with the true values being sensitive to, for example, the `discretization' of the template grids used in the fitting procedure. The systematic uncertainties on stellar masses and far infrared luminosities are likely to be around 0.3\,dex \citep{Gruppioni08,Mancini11,Santini15}. As $L_{\rm IR,SF}$ is important for our interpretation we also tested how reliable our values are for the targets that lack far infrared photometric detections, by re-fitting all of the other SEDs (with good FIR coverage) but removing the longer wavelength data. Through this exercise we find that even with no data at $\ge$22$\mu$m, the new values vary by no more than $\sim 2 \sigma$ from the values derived using all the available \emph{FIR} data\footnote{$\sigma$ is the formal error recorded in Table~\ref{SED_vals}.}. This is likely due to the additional constraints on the star formation that come from the optical part of the SEDs in our fitting procedure. These boosted (2$\sigma$) errors also make the values consistent with the $L_{\rm IR,SF}$ values presented for the same sources in \citet{Harrison14}. We flag these affected sources in Table~\ref{SED_vals} and show the $2 \sigma$ error bars in Fig.~\ref{radio_loudness}. We note that these additional systematic uncertainties do not affect the conclusions drawn in this paper. 

To estimate the star formation rates of our targets we used the SED-derived $L_{\rm IR,SF}$ and the relationship from \citet{Kennicutt12}, correcting to a Chabrier IMF by dividing by 1.7 \citep{Chabrier03}. In Table~\ref{SED_vals} the quoted uncertainties are only from the SED-derived $L_{\rm IR,SF}$ uncertainties. We note that there is an additional systematic uncertainty on the star formation rates due to the conversion factor from the far-infrared luminosity \citep[i.e., $\sim$0.3\,dex;][]{Kennicutt12}. 

Our targets are type~2 AGN, and consequently, we do not detect the primary AGN disc emission in the \emph{UV}--optical part of the SED. Therefore, to estimate the bolometric AGN luminosity ($L_{\rm AGN}$) we converted the 6$\mu$m luminosity from the AGN emission component using a bolometric correction of 8$\times$ following \citet{Richards06}. The uncertainties on these $L_{\rm AGN}$ values are dominated by a $\sim$1\,dex systematic uncertainty on the bolometric corrections. This confirms that our targets are consistent with having quasar luminosities (i.e., $L_{\rm AGN}$$\gtrsim$$10^{45}$\,erg\,s$^{-1}$)\footnote{We note that the bolometric luminosity calculated in this way is below the quasar limit for J1316$+$1753, however it is consistent with being equal to or above 10$^{45}$\,erg\,s$^{-1}$ within the 1~dex error mentioned above. Additionally, it is clearly a quasar using the [O~{\sc iii}] luminosity \citep{Reyes+08}.}, which we also find by using $L_{\rm [O III]}$ and the bolometric correction from \citet{Heckman04} (see Table~\ref{SED_vals}).


	\begin{table*}
	\caption{Galaxy and AGN parameters derived from SED fitting for the primary sample. \newline
    Notes: All values are given with 1$\sigma$ formal errors from the SED fit (see Section~\ref{characterising} for systematic uncertainties). (1) Object name; (2) Bolometric AGN luminosity; (3) Stellar mass; (4) Infrared luminosity from star formation in the range 8--1000 $\mu$m; (5) Star formation rate; (6) 1.4~GHz flux predicted from star formation following the radio -- $L_{IR}$ relation \citep{Bell+03}; (7) Percentage of the FIRST luminosity accounted for by star formation in the radio excess sources; (8) The $q_{\text{\emph{IR}}}$ (`radio excess') parameter, where $q_{\text{\emph{IR}}} \leq 1.8$ denotes radio excess (see Section~\ref{characterising}); (9) Flag to define if the target is radio excess, where: `Y' means radio excess, `P' means probably radio excess and `N' means not radio excess.
$^{\dagger}$These sources do not have photometric measurements at wavelengths longer than 60$\mu$m, with J1316$+$1753 having no photometry above 22$\mu$m (see Section~\ref{characterising} for a discussion on the additional uncertainties on the parameters for these sources). $^{\dagger\dagger}$ For this target the AGN contribution is particularly high in the \emph{NIR} regime and the estimate of the stellar mass is unconstrained, with an uncertainty larger than the parameter value itself. We therefore do not report a value of M$_{\star}$.
}   
	\centering  
	\begin{tabular}{llrrrccrc}
  \hline
  
 \multicolumn{1}{c}{ Name }&\multicolumn{1}{c}{ $\log[L_{\text{AGN}}]$ }& \multicolumn{1}{c}{ log[M$_{\star}$] }& \multicolumn{1}{c}{ $\log[L_{\text{IR,SF}}]$ }& \multicolumn{1}{c}{ SFR }& \multicolumn{1}{c}{ S$_{1.4,\text{SF}}^{\text{predicted}}$ }&\multicolumn{1}{c}{   \% SF } & \multicolumn{1}{c}{ $q_{\text{\emph{IR}}}$}& Radio Excess\\
& \multicolumn{1}{c}{ (erg s$^{-1}$)} &\multicolumn{1}{c}{ (M$_\odot$) }&\multicolumn{1}{c}{ (erg s$^{-1}$)} & \multicolumn{1}{c}{ (M$_\odot$yr$^{-1}$) }& \multicolumn{1}{c}{(mJy)} & &  \\
 \multicolumn{1}{c}{ (1) }& \multicolumn{1}{c}{(2)} & \multicolumn{1}{c}{ (3) }& \multicolumn{1}{c}{(4)} & \multicolumn{1}{c}{(5) }& \multicolumn{1}{c}{(6)}& \multicolumn{1}{c}{ (7) }& \multicolumn{1}{c}{(8)} & \multicolumn{1}{c}{(9)}\\ 
\hline
J0945+1737 & 45.7 & 10.1$^{+0.09}_{-0.12}$ & 45.3$\pm$0.02 & 46$\pm$2 & 3.0$\pm$0.2 & 6.9$\pm$0.4 & 1.6$\pm$0.02 & Y \\
J0958+1439$^{\dagger}$ & 45.2 & 10.74$^{+0.09}_{-0.12}$ & 44.6$^{+0.2}_{-0.3}$ & 10$\pm$5 & 0.9$\pm$0.5 & 9$\pm$4 & 1.7$^{+0.2}_{-0.5}$ & P \\
J1000+1242$^{\dagger}$ & 45.3 & 9.9$^{+0.3}_{-0.7}$ & 45.0$^{+0.1}_{-0.2}$ & 24$\pm$7 & 1.1$\pm$0.4 & 4$\pm$1 & 1.3$^{+0.2}_{-0.4}$ & Y \\
J1010+1413$^{\dagger}$ & 46.2 & 11.0$\pm$0.1 & 45.1$^{+0.2}_{-0.4}$ & 30$\pm$20 & 0.8$\pm$0.5 & 9$\pm$5 & 1.8$^{+0.3}_{-0.7}$ & P \\
J1010+0612 & 45.3 & 10.5$^{+0.3}_{-0.9}$ & 44.99$\pm$0.04 & 22$\pm$2 & 2.6$\pm$0.2 & 2.6$\pm$0.2 & 1.15$\pm$0.04 & Y \\
J1100+0846 & 46.0 & --$^{\dagger\dagger}$ & 45.01$^{+0.08}_{-0.09}$ & 24$\pm$5 & 2.6$\pm$0.5 & 4.3$\pm$0.8 & 1.37$^{+0.08}_{-0.09}$ & Y \\
J1316+1753$^{\dagger}$ & 44.4 & 11.0$^{+0.2}_{-0.3}$ & 45.1$^{+0.2}_{-0.3}$ & 30$\pm$10 & 1.3$\pm$0.6 & 11$\pm$6 & 1.8$^{+0.2}_{-0.5}$ & P \\
J1338+1503$^{\dagger}$ & 45.7 & 10.6$^{+0.1}_{-0.2}$ & 45.1$^{+0.2}_{-0.4}$ & 30$\pm$20 & 0.8$\pm$0.5 & -- & 2.3$^{+0.3}_{-0.7}$ & N \\
J1356+1026 & 45.2 & 10.64$^{+0.09}_{-0.11}$ & 45.36$\pm$0.02 & 53$\pm$3 & 3.8$\pm$0.2 & 6.4$\pm$0.3 & 1.56$\pm$0.02 & Y \\
J1430+1339 & 45.5 & 10.86$^{+0.05}_{-0.06}$ & 44.32$^{+0.06}_{-0.07}$ & 4.8$\pm$0.7 & 0.8$\pm$0.1 & 2.9$\pm$0.4 & 1.18$^{+0.06}_{-0.07}$ & Y \\
\hline
\end{tabular}

 \label{SED_vals} 
	\end{table*}

%

\begin{figure}
 \centering
 \includegraphics[width=\hsize]{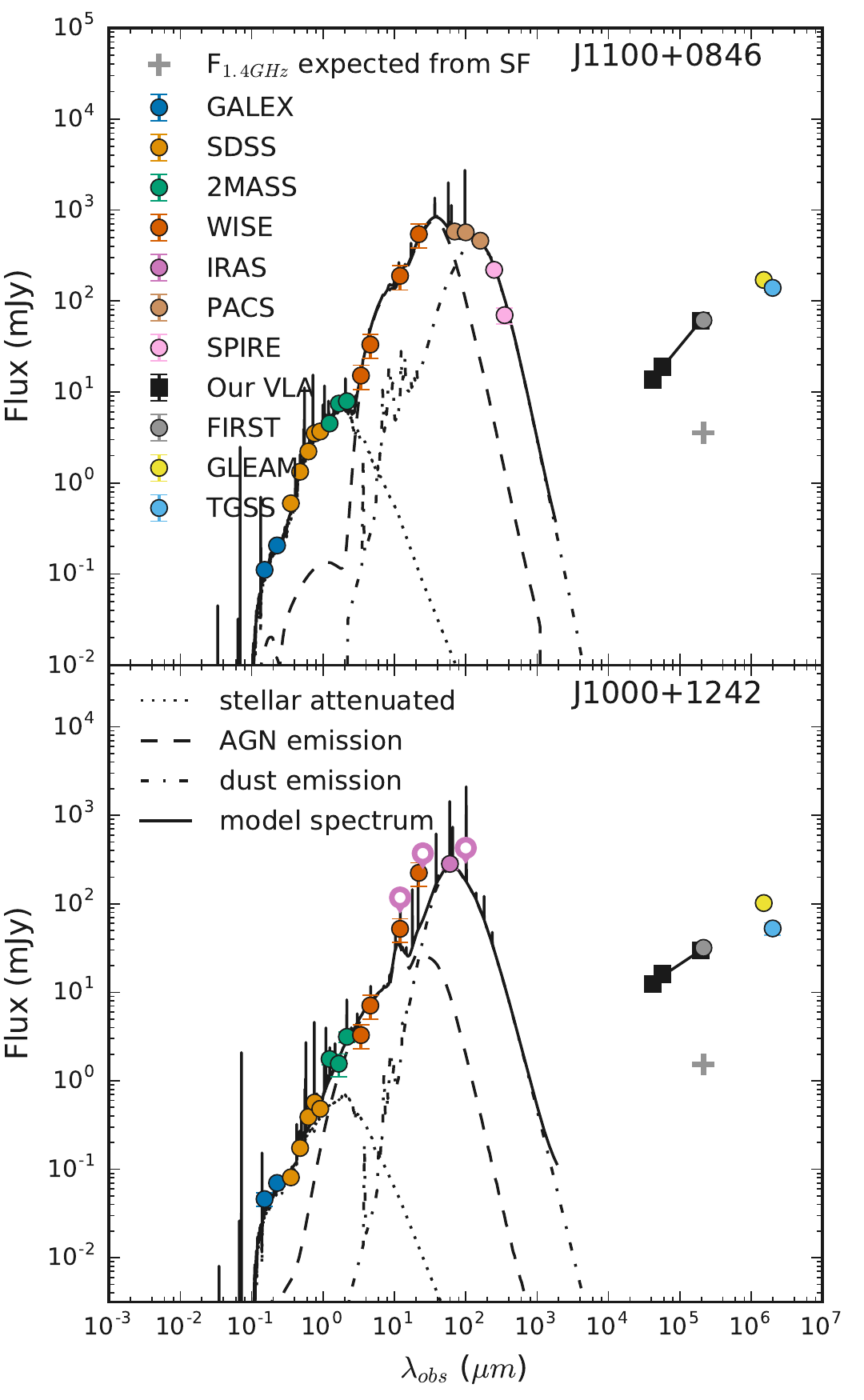}
 \caption{Two of the SEDs of our targets (J1100$+$0846; an example with good \emph{FIR} photometry and J1000$+$1242; an example without \emph{FIR} detections). The other SEDs are included in the supplementary online material. The data are plotted as circles colour coded by survey, with upper limits as open circles. The available radio photometry are shown but are not included in the fit. The solid curve shows our best total fit that is decomposed into attenuated stellar (dotted line), AGN (dashed line), and star-formation heated dust (dot-dashed line) components. An additional nebular component, not plotted separately, is responsible for the emission-lines visible in the total fit. A grey cross marks the predicted 1.4\,GHz flux from star formation following \citet{Bell+03}. We find that all but one of targets have excess radio emission above that expected from star formation (Section~\ref{characterising}).
  }
  \label{SED}
 \end{figure}

 Two example SEDs are shown in Fig.~\ref{SED} and the remainder are provided in online supplementary material. Although the SEDs are only fit using the \emph{UV}--\emph{FIR} photometry we also show the radio fluxes from three radio surveys, when available: FIRST (1.4~GHz); TIFR GMRT Sky Survey Alternative Data release \citep[TGSS; 150MHz; ][]{2017A&A...598A..78I}; and the GaLactic and Extragalactic All-sky MWA Survey \citep[GLEAM; 200MHz;][]{2017MNRAS.464.1146H}, as well as the total flux from our measurements at 1.5, 5.2 and 7.2~GHz (see Section~\ref{radio_SED} and Table~\ref{alphas}). These additional radio data allow for visual comparison to the level of emission expected from star formation alone \citep{Bell+03} and the identification of spectral turnovers in the radio SEDs (Section~\ref{jets}). 

%


In Fig.~\ref{radio_loudness} we show $L_{IR,SF}$ as a function of radio luminosity (see Table~\ref{targets}) for our targets and compare these values to the \emph{FIR}--radio correlation of star-forming galaxies from \citet{Bell+03}. It can be seen that nine of our ten targets lie well above the \emph{FIR}--radio correlation\footnote{If we instead used the \emph{FIR}--radio correlation from \citet{Delhaize17} the nine targets lie even further above the relationship of star-forming galaxies.} (the exception is J1338$+$1503). We note that for three of the targets, which all have poor \emph{FIR} photometry, the uncertainties could cause them to be consistent with star-forming galaxies, within the 3$\sigma$ scatter on that relation, and we highlight these targets in Table~\ref{SED_vals}. However, based on our spatially-resolved radio images presented in Section~\ref{VLA_imaging}, we are confident that these sources have significant radio emission that is not associated with star formation. The distance of points from this relation can be quantified using the $q_{\text{IR}}$ parameter, defined as 
\begin{equation} q_{\text{\emph{IR}}}= \log \left[ \frac{L_{\text{\emph{IR}}}/3.75 \times 10^{12} \text{W} }{L_{1.4} / \text{ W Hz}^{-1}} \right], \end{equation}
where $L_{IR}$ is the rest-frame far-infrared (8--1000 $\mu$m) luminosity, and sources with $q_{\text{\emph{IR}}}\lesssim1.8$ are considered radio excess \citep[see Table~\ref{SED_vals}; e.g.][]{Bell+03,DelMoro13,Delhaize17}.

In Table~\ref{SED_vals} we also show the radio flux that we expect from star formation following \citet{Bell+03} and we find that for our nine `radio excess' targets, the star formation is expected to contribute $\sim$3--11~per cent of the total observed radio emission (also see Fig.~\ref{SED}). We discuss this in more detail in Section~\ref{sec:star_formation}.

\section{Observations and data reduction}
\label{obs&redux}
\subsection{VLA observations, and imaging}
\label{VLA}

\subsubsection{Observations and data reduction}
We observed with VLA under two proposals: programme 13B-127, with observations carried out 2013 December 1-- 2014 May 13 and programme 16A-182 with observations carried out 2016 May 30 -- 2017 January 20. For 13B-127 we observed nine targets, from our primary sample of ten, in four configuration--frequency combinations: (1) A-array in L-band (1--2\,GHz; $\sim$1.3~arcsec\,resolution); (2) A-array in C-band (4--8\,GHz; $\sim$0.3~arcsec\,resolution); (3) B-array in L-band (1--2\,GHz; $\sim$4.3~arcsec\,resolution) and (4) B-array in C-band (4--8\,GHz; $\sim$1.0~arcsec\,resolution). The final target in our primary sample (J1338$+$1503)\footnote{along with two other targets from the \citet{Harrison14} sample not included in the primary sample (J1355$+$1300 and J1504$+$0151; see supplementary information)}, was observed by VLA during our 16A-182 project. Due to incomplete observations, this was only observed in one configuration--frequency combination: B-array configuration in the C-band (i.e., 4--8\,GHz; $\sim$1.0~arcsec\,resolution).

The 13B-127 observations comprise 2\,hours ($\sim$5 minutes on each target) of L-band observations in the A-configuration, and 2.5\,hours ($\sim$7--10 minutes on each target) of L-band observations in the B-configuration. For the C-band, 2.5\,hours ($\sim$7 minutes on each target) and 2\,hours ($\sim$5 minutes on each target) of observations were taken in A- and B-configurations, respectively. During our 16A-182 observations C-band data in B-configuration were taken of three targets. These were taken with 2$\times$ or 3$\times$ repeats of 1\,hour observing blocks (with 35, 32 and 26\,minutes on target for each J1338$+$1503; J1355$+$1300 and J1504$+$0151, respectively).

We perform amplitude and bandpass calibration at L- and C-band at the start of each observing block using a $\sim$10\,minute scan on the standard calibration source J1331+3030 (3C\,286), and determine complex gain solutions via $\sim$3\,minute scans (including slew-time) of nearby calibration sources every $\sim 10$--$15$\,minutes (typically within $\sim$10$^\circ$ of our targets). We choose calibration sources from the list of VLA calibrators website\footnote{\url{https://science.nrao.edu/facilities/vla/observing/callist}} with codes which deem them suitable for each combination of array configuration/observing frequency. The 13B-127 observations were reduced using the Common Astronomy Software Applications ({\sc casa}\footnote{\url{https://casa.nrao.edu/}}) package (version 4.1.0), along with version 1.2.0 of the VLA scripted pipeline. We reduced the later 16A-182 observations using {\sc casa} version 4.5.2, and version 1.3.5 of the VLA scripted pipeline.

\begin{table*}

\caption{Summary of the radio images used in Fig.~\ref{radio_collage_dat}. \newline
Notes: (1) object name; (2) resolution of the image; (3-5) details of the synthesised beams and noise of the radio images; (6) describes the measurement set(s) used, where C-A indicates the VLA C-band A-configuration data (etc.); (7) describes the weighting scheme used to image the data. $^{\dagger}$a concatenation of the C-band A and B-array data were used with relative weighting of 4:1; $^{\dagger\dagger}$a concatenation of the C-band A and B-array data were used with even weight.
} 

\centering    

\begin{tabular}{c c c d{2.0} d{3.0} c c } 
\hline  
Object Name & Image & \multicolumn{1}{c}{Beam HPBW} & \multicolumn{1}{c}{Beam PA} & \multicolumn{1}{c}{Noise} & Data & Weighting \\ 
 & & \multicolumn{1}{c}{(arcsec)} & \multicolumn{1}{c}{ (deg) }& \multicolumn{1}{c}{($\mu$Jy/beam) } \\
 (1) & (2) & (3) & (4) & (5) & (6) & (7) \\
\hline   

J0945$+$1737 & HR & 0.22$\times$0.21 & -18 & 19 & C-A & uniform \\
J0945$+$1737 & LR & 1.17$\times$0.94 & 52 & 18 & C-B & briggs 0.5 \\
J0958$+$1439 & HR & 0.22$\times$0.21 & -27 & 12 & C-A & uniform \\
J0958$+$1439 & LR & 1.22$\times$0.94 & 51 & 9 & C-B & briggs 0.5 \\
J1000$+$1242 & HR & 0.30$\times$0.26 & -12 & 23 & C-A & briggs 0.5 \\
J1000$+$1242 & LR & 1.44$\times$1.21 & 39 & 22 & C-A+C-B$^{\dagger}$ & natural \& 150k$\lambda$ taper \\
J1010$+$1413 & HR & 0.22$\times$0.21 & -41 & 16 & C-A & uniform \\
J1010$+$1413 & LR & 1.21$\times$0.95 & 51 & 9 & C-B & briggs 0.5 \\
J1010$+$0612 & HR & 0.25$\times$0.22 & -68 & 79 & C-A & uniform \\
J1010$+$0612 & LR & 1.55$\times$0.93 & 50 & 83 & C-B & briggs 0.5 \\
J1100$+$0846 & HR & 0.23$\times$0.22 & -67 & 48 & C-A & uniform \\
J1100$+$0846 & LR & 1.24$\times$0.94 & 48 & 58 & C-B & briggs 0.5 \\
J1316$+$1753 & HR & 0.26$\times$0.22 & 78 & 14 & C-A & uniform \\
J1316$+$1753 & LR & 1.01$\times$0.86 & 29 & 16 & C-B & briggs 0.5 \\
J1338$+$1503 & LR & 1.01$\times$0.91 & 51 & 6 & C-B & briggs 0.5 \\
J1356$+$1026 & HR & 0.37$\times$0.29 & -47 & 100 & C-A & briggs 0.5 \\
J1356$+$1026 & LR & 1.05$\times$0.92 & 4 & 36 & C-A+C-B$^{\dagger\dagger}$ & natural \\
J1430$+$1339 & HR & 0.33$\times$0.23 & -88 & 23 & C-A & uniform \\
J1430$+$1339 & LR & 1.00$\times$0.88 & 16 & 12 & C-B & briggs 0.5 \\

\hline     
\end{tabular}

\label{radio_collage_tbl}
\end{table*}


\subsubsection{Imaging }
\label{VLA_imaging}

When imaging the VLA data we had two goals: (1) identify any morphological features that are present in the radio emission and (2) measure the fluxes / spectral indices of these features. These goals required different approaches to the imaging of the data and we describe them both here. In order to take into account the broad and varying bandwidths of our observations, all of the VLA images we present were made using the Multi-Frequency Synthesis (MFS) mode of the {\sc clean} function in {\sc casa} version 4.7.1. We chose the weighting of the baselines to obtain the desired compromise between sensitivity and beam size to achieve our science goals (see \citealt{Briggs95}). In some cases we additionally applied Gaussian tapering to achieve the desired resolution. Because of the relatively short observing times of our targets and resultant limited $uv$ coverage our cleaned images sometimes still suffer from relatively strong beam residuals. However, in order to test the validity of these features we ensured that they were identified at multiple frequencies and using different combinations of weighting and tapering (see Section~\ref{radio_SED}).

\begin{figure*}

 \includegraphics[width=16.5cm]{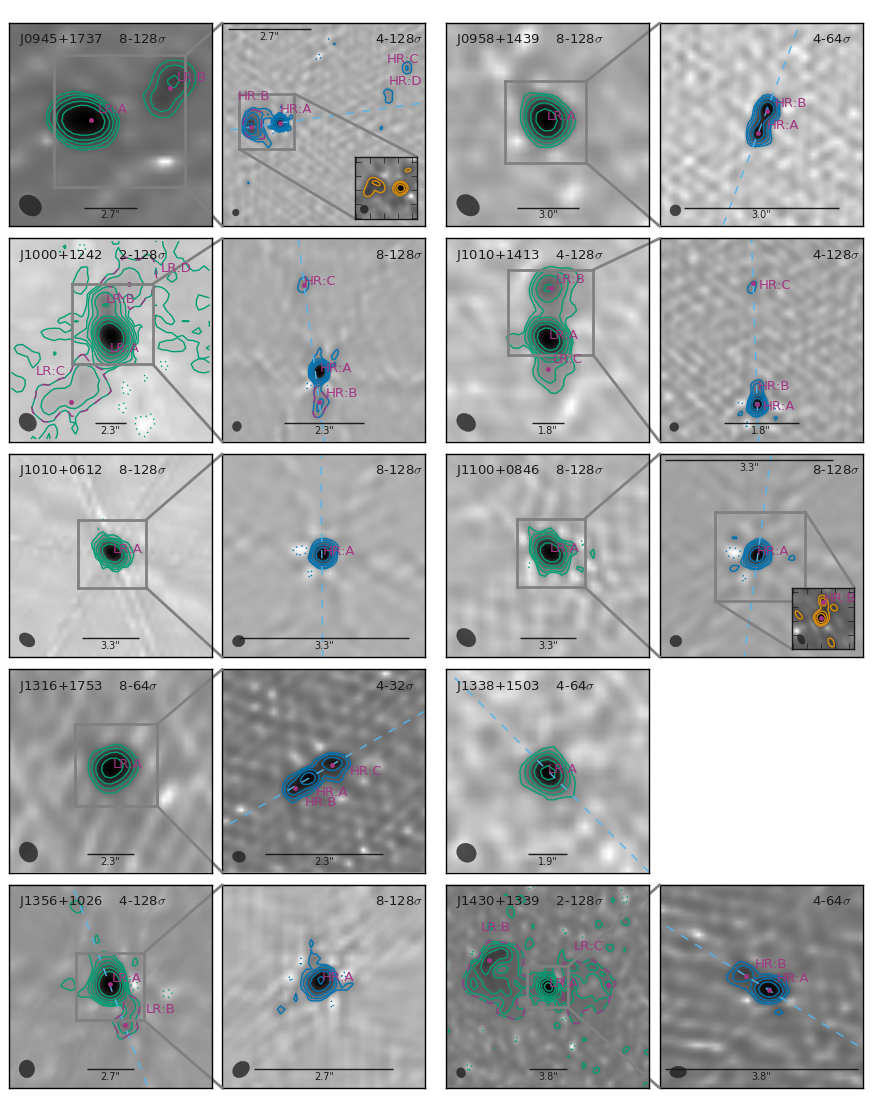}
 \centering
 \caption{VLA 6~GHz images and contours for our primary sample. The low-resolution images ($\sim$1\,arcsec beam; green contours) are shown to the left and on the right are the high-resolution images ($\sim$0.25\,arcsec beam; blue contours) that cover the region marked with grey boxes in the low-resolution images. For J0945$+$1737 and J1100$+$0846, we also inset the 1.5\,GHz e-MERLIN images (orange contours). Contours are at $\pm[2, 4, 8, 16, 32, 64, 128]\sigma$ with the minimum and maximum contour in each image given in the top right (see Section~\ref{VLA_imaging}; the e-MERLIN contours always start at 8$\sigma$). The synthesised beams are shown as black ellipses and the scale bars represent 6\,kpc. We label the radio morphological features and when a region is used to calculate the flux it is shown by a dashed magenta outline (Section~\ref{radio_SED}). Magenta dots show the positions used to calculate sizes, distances, and the major axis of the radio emission (light-blue dashed line ; see Section~\ref{morph}). In seven sources we see unambiguous multiple radio features. Additionally, J1100$+$0846 shows one ambiguous feature, only seen in the e-MERLIN image. Only for J1338$+$1503, which lacks a high-resolution image, and for J1010$+$0612 do we see no evidence of multiple radio features.}
  \label{radio_collage_dat} 
  
 \end{figure*}

For our first goal, we aimed to identify both diffuse and compact morphological features in the radio emission. To do this we made two C-band (6\,GHz) `showcase images' for each galaxy in the primary sample: one with a $\sim$1~arcsec beam (i.e., $\sim$2~kpc at $z$=$0.1$; referred to as `low-resolution' or LR) and one with a $\sim$0.25~arcsec beam (i.e., $\sim$0.5~kpc at $z$=$0.1$; referred to as `high-resolution' or HR)\footnote{For J1338$+$1503 only a low-resolution image was created due to there being no available A-configuration VLA data.}. Fig.~\ref{radio_collage_dat} shows the low and high-resolution images for each source. For these showcase images, the imaging parameters (weighting, tapering and concatenating data sets) were tweaked to best reveal the morphological features found in each galaxy (details in Table~\ref{radio_collage_tbl}). The $\sigma$ (noise) values are calculated using 8$\sigma$ clipping repeated ten times across a region of the images that is 50 times the size of the beam major axis. 

For our second goal we measured the flux densities and spectral indices (over 1-7GHz) for each morphological feature identified. To do this we required multi-frequency images with the same spatial resolution. We produced images at 1.5, 5.2 and 7.2~GHz, where the 5.2 and 7.2~GHz images were made by evenly splitting the 16 VLA C-band spectral windows. We created two sets of these resolution-matched multi-frequency images using weighting and tapering to match the beams as closely as possible, one set at high-resolution ($\sim$0.25~arcsec; using e-MERLIN for the 1.5~GHz data; see Section~\ref{e-MERLIN}) and one set at low-resolution ($\sim$1~arcsec; using the VLA L-band A-configuration data for the 1.5~GHz images). Full details of the data used, applied weighting schemes/tapering, the resultant properties of the images (i.e., beam sizes, noise levels; see Table~B1) and all 56 of these images, are presented in the online supplementary material.  

For the two sources observed with our VLA 16A-182 programme but not included in the primary sample (J1355$+$1300 and J1504$+$0151), the details of the imaging, the images themselves and a brief discussion of the features seen are presented in the online supplementary material (Appendix C).

\subsection{e-MERLIN observations and imaging}
\label{e-MERLIN}

We obtained 1.5\,GHz (L-band) e-MERLIN observations with $\sim$40--172\,minutes on-source per target in Cycle 1 (ID: CY\,1022; observed on 2013 December 19) and $\sim$1.5--3\,hours on-source per target in Cycle 2 (ID: CY\,2217; observed between 2015 January 21--23). Each observing block entailed $\sim$30\,minute scans of $0319$+$415$ and $1407$+$284$ for flux density and bandpass calibration, respectively. Target scans were $\sim$8\,minutes, interspersed with $\sim$3\,minute scans of bright, nearby phase reference sources. We processed the data in {\sc aips} version {\sc 31dec15}, with most steps being carried out using the e-MERLIN pipeline \citep{argo15}, but with extensive additional manual flagging of bad data using the {\sc aips} tasks {\sc spflg} and {\sc ibled}. 

We created $4096\times 4096$\,pixel maps from our calibrated $uv$ data using {\sc imagr} with a pixel size of 0.05\,arcsec (i.e. a 3\,arcminute postage stamp around each source), and iteratively deconvolved the point spread function (PSF) within manually-defined {\sc clean} boxes around sources of bright emission. Our typical synthesised beam is $\sim$0.2--0.3\,arcsec and in all cases the beams have axial ratios of $b/a > 0.5$. 

Unfortunately, a combination of strong, persistent radio frequency interference (RFI) and hardware failures resulted in the loss of $>$50~per cent of our Cycle 1 data, covering both target and calibrator fields. The resulting loss of point-source sensitivity and $uv$ coverage severely limited the quality of the final images, leaving us unable to achieve our science goals of establishing the radio morphologies at 1.5\,GHz. Consequently we do not use these data in our analyses. Fortunately, only one of our primary targets (J1338$+$1503) was part of this Cycle 1 programme. Improvements in both the hardware performance and observing strategies for Cycle 2 lowered our loss-rate to $\sim$20~per cent (with significantly improved $uv$ coverage), yielding improved imaging ($\sigma\approx 300\,\mu$Jy\,beam$^{-1}$) and allowing us to study the 1.5\,GHz morphologies on sub-kpc scales of the nine of our primary targets, that were all observed in this cycle.

\subsection{IFS observations and data reduction}
\label{IFS}
All of our quasars have published IFS data obtained using Gemini-GMOS \citep{Harrison14}. These observations covered the O~[{\sc iii}]$\lambda$4959, 5007 and H$\beta$ emission-lines using 25 $\times$ 20 lenslets sampling a 5 $\times$ 3.5~~arcsec field of view. The spectral resolution of $\sim$3700 gives a line FWHM of 80~km s$^{-1}$. The observations were performed with a typical V band seeing of $\sim$ 0.7~arcsec. More details about these observations are given in \citet{Harrison14}.

As part of an uncompleted ESO programme, three of our targets (J1430$+$1339, J1010$+$1413 and J1000$+$1242) also have IFS observations with VIMOS on the ESO/VLT telescope observed from 2014 January 23--24 and 2014 March 9--10 (Program ID: 092.B-0062). The VIMOS observations, which benefit from a $\sim$20$\times$20~arcsec field of view, were motivated by the kinematic and morphological structures that appeared to extend beyond the GMOS field of view \citep[i.e., on $\gtrsim$5~arcsec scales;][ see Fig.~\ref{sdss}]{Harrison14}. The data for J1430$+$1339 are already published in \citet{Harrison15} and we combine these data with the rest of the sample here. 

We used VIMOS in IFS mode, using the HR-Orange grism, which provides a wavelength range of 5250--7400\AA\, at a spectral resolution of $\sim$2650, giving a line FWHM of 110\,km s$^{-1}$ at 5007\AA, which we confirmed within $\pm$10\,km s$^{-1}$ by measurements of sky-lines. During the observations the targets were dithered around the four quadrants of the VIMOS field of view. The on-source exposure times were 6480\,seconds for J1430$+$1339 and J1010$+$1413; and 2160\,seconds for J1000$+$1242. The V-band seeing ranged between 0.8--0.9~arcsec. Standard stars were taken under similar conditions to the science observations. The standard {\sc esorex} pipeline was used to reduce the data, which includes bias subtraction, flat-fielding, wavelength calibration and flux calibration. Data cubes were constructed from the individually sky-subtracted, reduced science frames. The final data cubes were created by median combining the individual exposure cubes using a three-sigma clipping threshold.

\section{Analyses}
\label{analysis}

Here we describe the techniques used to identify and characterise morphological and kinematic features observed in our radio and optical IFS data. In Section~\ref{radio_analysis} we identify the radio features seen at different resolution and measure the location, flux and spectral index for each feature. In Section~\ref{line_fitting}, we describe the non-parametric characterisation of the [O~{\sc iii}]$\lambda 5007$ emission-line profiles and explain how we produced emission-line kinematic maps.

\subsection{Radio analyses}
\label{radio_analysis}

\subsubsection{Radio features, flux densities and spectral indices}
\label{radio_SED}

As can be seen in Fig.~\ref{radio_collage_dat} our sources are typically composed of multiple spatially-distinct radio features. In order to constrain the source of the radio emission in each of these morphological features (discussed in Section~\ref{sec:origin}), we calculated flux densities and spectral indices for each. We note that we tested our overall approach to flux calibration and to obtaining flux densities by verifying that the total fluxes of the sources from our imaged VLA L-band B-array data (average spatial resolution of 4.3~arcsec) are consistent with the FIRST values (5~arcsec beam) within errors. We note that the variations in flux densities by converting between the difference in the central frequency of our observations and FIRST (i.e., 1.5\,GHz to 1.4~GHz) is smaller than the errors on our fluxes.

We name each morphologically distinct feature (detected at $\ge$3$\,\sigma$) in the high-resolution images as HR:A, HR:B etc., and similarly for the low-resolution images with LR:A, LR:B etc. For the low significance features (e.g.\ HR:C in J1000$+$1242) we verified that they are real by ensuring that they were significantly detected in images produced using multiple weighting schemes and/or in independent observations. In general the e-MERLIN images (see Section~\ref{e-MERLIN}) did not reveal any new information on the morphological features. The exceptions are: for J0945$+$1737, where the HR:B feature shows a bent `jet like' appearance in the e-MERLIN image; and for J1100$+$0846, which shows a $\sim$7$\sigma$ feature (HR:B; see Table~\ref{alphas}) in the e-MERLIN image that is not identified in the 6~GHz VLA images. We show the e-MERLIN images for these two sources in Fig.~\ref{radio_collage_dat}. The e-MERLIN images for all sources are presented in the supplementary online material. We discuss the origin of the identified radio features in Section~\ref{sec:origin}.

Due to the range of morphologies seen in our data (see Fig.~\ref{radio_collage_dat}), we were required to use two approaches to obtain the flux densities of each feature. The first approach was to model the emission as a series of two-dimensional Gaussian components. All of the parameters of the fits were left free\footnote{For the LR-C component of J1010$+$1413 we were required to fix the peak position of the Gaussians to within 0.5~pixels to obtain a reasonable fit. We flag this feature as having unreliable flux density measurements.}. We note that the feature LR:A in both J0945$+$1737 and J1000$+$1242 needed two component Gaussians to provide an adequate fit, which is easily explained by the multiple HR components that they are composed of (see Fig.~\ref{radio_collage_dat}). 

The second approach was to sum the emission in regions motivated by the lowest-level contours of the appropriate resolution image in Fig.~\ref{radio_collage_dat}. We verified that the flux density measurements are consistent within the errors (described below) if we vary the defined region sizes up or down by 25~per cent. These regions were primarily used for diffuse/irregular structures that are not well described by a Gaussian and are referred to as `region components'. Where it was possible to apply both approaches, we further verified that they gave consistent results, but favoured the Gaussian fitting method. In the cases where a compact nuclear component was seen in addition to a more diffuse structure, the flux in the diffuse regions was calculated after subtracting off the Gaussian fits to the compact component(s) in order to minimize contamination. Figures showing the data, our best-fit models and the corresponding residuals for all multi-frequency images can be found in the online supplementary material.

To calculate the random noise on our flux density measurements for each feature, we took the standard deviation of 100$\times$ repeats of extracting flux densities from inside appropriately sized regions randomly positioned within the central 10~arcsec (for high-resolution images) or 20~arcsec (for the low-resolution images) avoiding source emission. For the `region components' the regions used in this procedure were the same size and shape as those used to extract the flux densities. For the Gaussian components we used an ellipse with axis sizes equal to twice the semi-minor and semi-major axes of the fits. To establish if a feature was detected in each of the 1.5, 5.2 and 7.2\,GHz images, we imposed a 5$\sigma$ detection limit. For each of the detected components, we added an additional 10~per cent systematic error, in quadrature, to the uncertainties to account for the random variations we found when extracting flux densities when changing the weighting scheme used to image the data. The final flux densities (or 5$\sigma$ upper limits) and their 1$\sigma$ uncertainties at 1.5, 5.2 and 7.2\,GHz are presented in Table~\ref{alphas}. 

\begin{figure}
 \centering
 \includegraphics[width=\hsize]{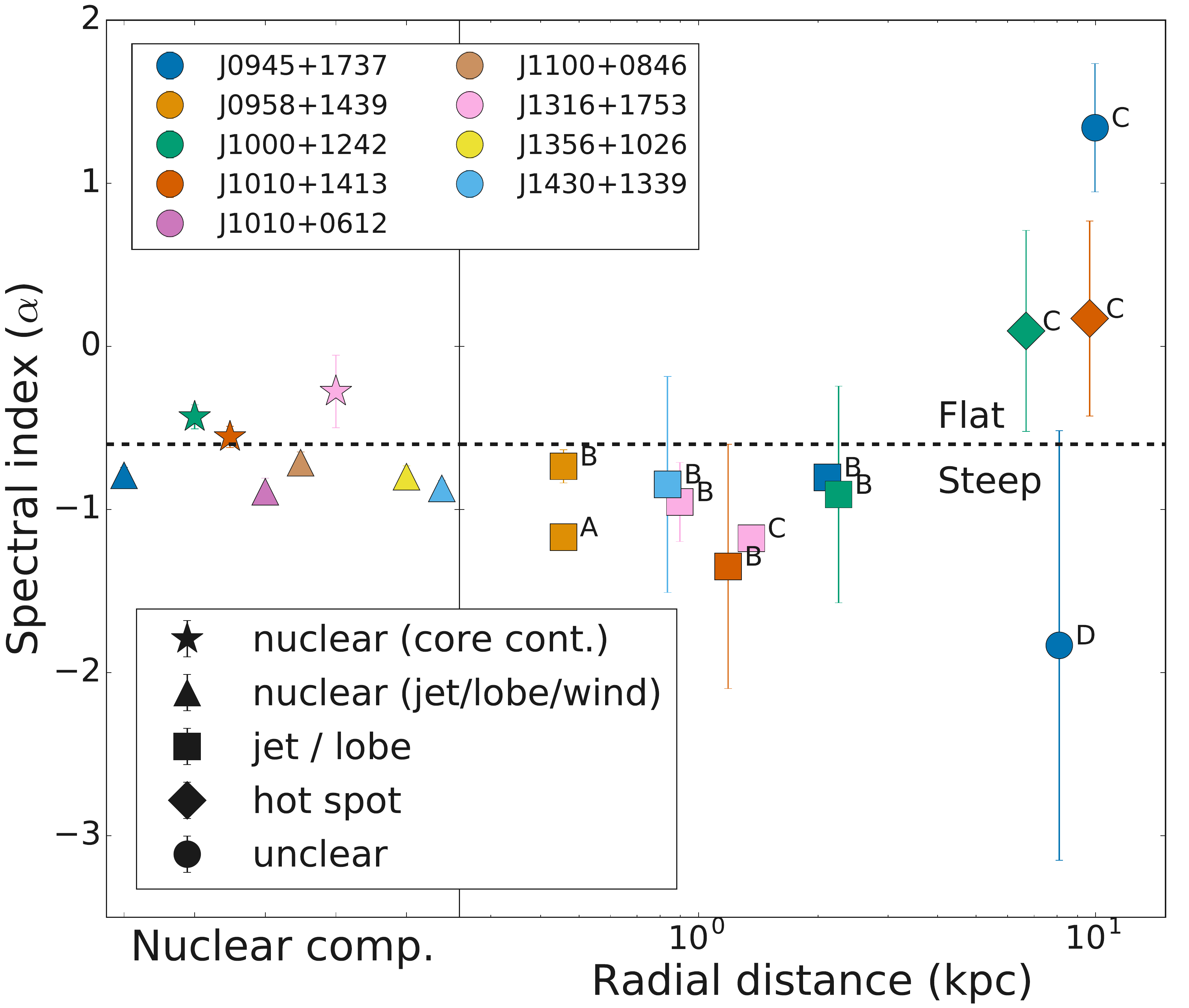}
 \caption{Spectral index ($\alpha$) of each high-resolution morphological radio component with detections in more than one band verses its distance to the brightest high-resolution radio component (i.e., nuclear/HR-A; Fig.~\ref{radio_collage_dat} ; Table~\ref{alphas}). The $\alpha$ values for the nuclear components are plotted on the left with an arbitrary separation. Each are colour coded based on their source, with the shape corresponding to our classification (see Section~\ref{sec:origin}). The nuclear components are noted as either jet/lobe/wind dominated or with a possible core contribution based on if they are steep ($\alpha<-0.6$) or flat ($\alpha>-0.6$; dashed line). Non-nuclear components are labelled as either `jet/lobe' or as `hot spot' depending on if they are steep or flat. Two components have an `unclear' origin and are discussed in more detail in Section~\ref{sec:origin}.
  }
  \label{alpha_vs_dist}
 \end{figure}

The spectral index ($\alpha$, which we define as $S_\nu \propto \nu^{\alpha}$) is used to help interpret the source of the radio emission for each feature in Section~\ref{sec:origin}. We measured this by fitting a line through all of the detected frequencies (1.5, 5.2 and 7.2 GHz) for each radio structure. The errors given for $\alpha$ in Table~\ref{alphas} are the 1$\sigma$ errors on the fit, except for where the component was only detected in two bands, then the error is the propagated error from the two measured fluxes. Some of our data were observed at different epochs (e.g., when combining our e-MERLIN and VLA C-band B-configuration data) which risks unknown variability of the fluxes affecting the spectral index values. However, we find that our $\alpha$ values are consistent to those calculated using only our single-epoch VLA C-band data (within the errors on the VLA C-band values). The 1-7\,GHz radio SEDs for each radio component are presented in the supplementary material. The spectral indices for the high-resolution components are plotted in Fig.~\ref{alpha_vs_dist} along with their distance from the central brightest component (HR:A; see Section~\ref{morph}).


%



%
\begin{table*}
\centering  
\caption{Radio properties of the morphological structures (see Fig.~\ref{radio_collage_dat}), extracted from resolution matched e-MERLIN and VLA images. \newline
Notes: (1) object name; (2) the flux density from the FIRST survey; (3) name of structure; (4) interpretation of structure (see Section~\ref{sec:classification}), or largest linear size observed at that resolution in kpc; (5)-(7) flux density in mJy at 1.5, 5.2 and 7.2 GHz. The errors given are a combination of 1$\sigma$ random errors and a 10~per cent systematic (see Section~\ref{radio_SED}). 5$\sigma$ upper limits are given for non-detections; (8) the spectral index ($\alpha$) defined as $S_\nu \propto \nu^{\alpha}$ and found by fitting a line through all detected points between 1.5--7.2GHz. The errors quoted are discussed in Section~\ref{radio_SED}. $^{\ddagger}$ Although this component is not detected at 5$\sigma$ significance in any individual band it is detected in all three at 3$\sigma$ resulting in a spectral index of -0.5. $^{\dagger}$ Due to the fitting constraints needed to get this component to be well fit by a Gaussian in all three images the fluxes and spectral index for this source are unreliable (see Section~\ref{radio_SED}). $^{\dagger \dagger }$ We determine that HR:B is either a high S/N beam artefact or a variable component, which then determines if LR:A is a composite or not (see Section~\ref{sec:classification}).
} 
\begin{tabular}{l l l l r r r r r } 
\hline
\multicolumn{1}{c}{Name} & \multicolumn{1}{c}{S$_{\text{FIRST}}$} & \multicolumn{1}{c}{Structure} & \multicolumn{1}{c}{Interpretation} & \multicolumn{1}{c}{$S_{1.5\text{GHz}}$ }& \multicolumn{1}{c}{$S_{5.2\text{GHz}}$} & \multicolumn{1}{c}{$S_{7.2\text{GHz}}$} & \multicolumn{1}{c}{ $\alpha$} \\ 
&\multicolumn{1}{c}{(mJy)} & & \multicolumn{1}{c}{or LLS} & \multicolumn{1}{c}{(mJy) }&\multicolumn{1}{c}{ (mJy)} &\multicolumn{1}{c}{ (mJy) }\\
\multicolumn{1}{c}{ (1)} & \multicolumn{1}{c}{ (2) }& \multicolumn{1}{c}{ (3) }&\multicolumn{1}{c}{ (4) } & \multicolumn{1}{c}{(5) }&\multicolumn{1}{c}{ (6)} & \multicolumn{1}{c}{(7)} & \multicolumn{1}{c}{(8)}\\
\hline   
J0945+1737 & 44.5$\pm$0.4 & HR:A & nuclear (jet/lobe/wind) & 16$\pm$2 & 6.3$\pm$0.6 & 4.6$\pm$0.5 & -0.79$\pm$0.05 \\
 &  & HR:B & jet / lobe & 13$\pm$2 & 4.8$\pm$0.5 & 3.9$\pm$0.4 & -0.8$\pm$0.02 \\
 &  & HR:C & unclear & $<$7.5 & 0.24$\pm$0.04 & 0.37$\pm$0.04 & 1.3$\pm$0.4 \\
 &  & HR:D & unclear & $<$1.6 & 0.26$\pm$0.04 & 0.15$\pm$0.03 & -2$\pm$1 \\
 &  & HR:Total & LLS=2.1kpc & 29$\pm$2 & 11.7$\pm$0.9 & 9.0$\pm$0.8 & -0.76$\pm$0.02 \\
 &  & LR:A & composite & 42$\pm$2 & 13$\pm$1 & 9$\pm$1 & -0.927$\pm$0.007 \\
 &  & LR:B & lobe & 4.5$\pm$0.5 & 1.9$\pm$0.2 & 1.4$\pm$0.2 & -0.74$\pm$0.03 \\
 &  & LR:Total & LLS=11kpc & 47$\pm$1 & 15$\pm$1 & 11$\pm$1 & -0.906$\pm$0.008 \\
J0958+1439 & 10.4$\pm$0.4 & HR:A & jet / lobe & 5.8$\pm$0.7 & 1.3$\pm$0.1 & 0.97$\pm$0.1 & -1.17$\pm$0.07 \\
 &  & HR:B & jet / lobe & 3.7$\pm$0.7 & 1.6$\pm$0.2 & 1.2$\pm$0.1 & -0.7$\pm$0.1 \\
 &  & HR:Total & LLS=0.9kpc & 10$\pm$1 & 2.9$\pm$0.5 & 2.1$\pm$0.4 & -0.959$\pm$0.001 \\
 &  & LR:A & composite & 11$\pm$1 & 2.9$\pm$0.3 & 2.1$\pm$0.2 & -1.057$\pm$0.005 \\
J1000+1242 & 31.8$\pm$0.4 & HR:A & nuclear (core cont.) & 20$\pm$2 & 13$\pm$1 & 9$\pm$1 & -0.43$\pm$0.07 \\
 &  & HR:B & jet / lobe & $<$2.1 & 1.0$\pm$0.1 & 0.73$\pm$0.08 & -0.9$\pm$0.7 \\
 &  & HR:C & hot spot & $<$1.1 & 0.65$\pm$0.08 & 0.7$\pm$0.1 & 0.1$\pm$0.6 \\
 &  & HR:Total & LLS=8.9kpc & 20$\pm$1 & 14$\pm$1 & 11$\pm$1 & -0.35$\pm$0.09 \\
 &  & LR:A & composite & 25$\pm$2 & 14$\pm$1 & 11$\pm$1 & -0.52$\pm$0.05 \\
 &  & LR:B & lobe & 3.1$\pm$0.3 & 1.2$\pm$0.1 & 1.0$\pm$0.1 & -0.742$\pm$0.006 \\
 &  & LR:C & lobe & 1.3$\pm$0.3 & 0.7$\pm$0.1 & 0.6$\pm$0.1 & -0.5$\pm$0.1 \\
  &  & LR:D$^{\ddagger}$ & lobe & $<$0.69 & $<$0.52 & $<$0.44 & -- \\
 &  & LR:Total & LLS=25kpc & 30$\pm$1 & 16$\pm$1 & 12$\pm$1 & -0.54$\pm$0.05 \\
J1010+1413 & 8.8$\pm$0.5 & HR:A & nuclear (core cont.) & 5.9$\pm$0.7 & 3.2$\pm$0.3 & 2.4$\pm$0.2 & -0.55$\pm$0.07 \\
 &  & HR:B & jet / lobe & $<$2.4 & 0.41$\pm$0.04 & 0.27$\pm$0.03 & -1.3$\pm$0.7 \\
 &  & HR:C & hot spot & $<$3.4 & 0.22$\pm$0.03 & 0.23$\pm$0.03 & 0.2$\pm$0.6 \\
 &  & HR:Total & LLS=9.7kpc & 5.9$\pm$0.8 & 3.8$\pm$0.6 & 2.9$\pm$0.5 & -0.43$\pm$0.1 \\
 &  & LR:A & composite & 7.5$\pm$0.8 & 3.7$\pm$0.4 & 2.7$\pm$0.3 & -0.63$\pm$0.06 \\
 &  & LR:B & lobe & 1.1$\pm$0.1 & 0.72$\pm$0.09 & 0.56$\pm$0.06 & -0.43$\pm$0.08 \\
 &  & LR:C$^{\dagger}$ & lobe & $<$0.48 & 0.2$\pm$0.04 & 0.27$\pm$0.04 & 0.9$\pm$0.6 \\
 &  & LR:Total & LLS=15kpc & 8.6$\pm$0.9 & 4.6$\pm$0.6 & 3.6$\pm$0.5 & -0.55$\pm$0.05 \\
J1010+0612 & 99.3$\pm$0.3 & HR:A & nuclear (jet/lobe/wind) & 80$\pm$10 & 30$\pm$3 & 20$\pm$2 & -0.89$\pm$0.08 \\
 &  & LR:A & -- & 97$\pm$10 & 29$\pm$3 & 20$\pm$2 & -1.01$\pm$0.05 \\
J1100+0846 & 61.3$\pm$0.3 & HR:A & nuclear (jet/lobe/wind) & 42$\pm$5 & 19$\pm$2 & 13$\pm$1 & -0.71$\pm$0.07 \\
 &  & HR:B & artefact / variable$^{\dagger\dagger}$ & 20$\pm$3 & $<$0.1 & $<$0.13 & -- \\
 &  & HR:Total & LLS=0.8kpc & 62$\pm$2 & 19$\pm$1 & 13$\pm$1 & -0.98$\pm$0.01 \\
 &  & LR:A & composite$^{\dagger\dagger}$ & 61$\pm$6 & 19$\pm$2 & 14$\pm$1 & -0.95$\pm$0.02 \\
J1316+1753 & 11.4$\pm$0.4 & HR:A & nuclear (core cont.) & 1.3$\pm$0.3 & 1.1$\pm$0.1 & 0.83$\pm$0.08 & -0.3$\pm$0.2 \\
 &  & HR:B & jet / lobe & 1.8$\pm$0.3 & 0.69$\pm$0.07 & 0.4$\pm$0.04 & -1.0$\pm$0.2 \\
 &  & HR:C & jet / lobe & 4.2$\pm$0.6 & 0.92$\pm$0.09 & 0.67$\pm$0.07 & -1.18$\pm$0.05 \\
 &  & HR:Total & LLS=1.4kpc & 7.2$\pm$0.9 & 2.7$\pm$0.4 & 1.9$\pm$0.3 & -0.85$\pm$0.08 \\
 &  & LR:A & composite & 11$\pm$1 & 3.0$\pm$0.3 & 2.2$\pm$0.2 & -1.019$\pm$0.001 \\
J1338+1503 & 2.4$\pm$0.4 & LR:A & star formation? & -- & 0.7$\pm$0.07 & 0.53$\pm$0.05 & -0.9$\pm$0.6 \\
J1356+1026 & 59.6$\pm$0.4 & HR:A & nuclear (jet/lobe/wind) & 49$\pm$5 & 19$\pm$2 & 14$\pm$1 & -0.8$\pm$0.07 \\
 &  & LR:A & -- & 58$\pm$6 & 20$\pm$2 & 15$\pm$1 & -0.88$\pm$0.03 \\
 &  & LR:B & unclear & $<$0.88 & 0.57$\pm$0.07 & 0.4$\pm$0.06 & -1.1$\pm$0.8 \\
 &  & LR:Total & LLS=5.6kpc & 58$\pm$2 & 21$\pm$1 & 15$\pm$1 & -0.86$\pm$0.03 \\
J1430+1339 & 26.4$\pm$0.4 & HR:A & nuclear (jet/lobe/wind) & 6.3$\pm$0.9 & 2.3$\pm$0.2 & 1.6$\pm$0.2 & -0.87$\pm$0.05 \\
 &  & HR:B & jet / lobe & $<$2.1 & 0.69$\pm$0.08 & 0.53$\pm$0.06 & -0.8$\pm$0.7 \\
 &  & HR:Total & LLS=0.8kpc & 6.3$\pm$0.9 & 2.9$\pm$0.5 & 2.1$\pm$0.4 & -0.69$\pm$0.1 \\
 &  & LR:A & composite & 12$\pm$1 & 3.4$\pm$0.3 & 2.4$\pm$0.2 & -1.02$\pm$0.01 \\
 &  & LR:B & lobe & 10$\pm$1 & 3.1$\pm$0.3 & 2.2$\pm$0.2 & -0.97$\pm$0.03 \\
 &  & LR:C & lobe & 1.8$\pm$0.2 & 0.7$\pm$0.08 & 0.59$\pm$0.07 & -0.72$\pm$0.03 \\
 &  & LR:Total & LLS=19kpc & 24$\pm$1 & 7.2$\pm$0.7 & 5.2$\pm$0.6 & -0.97$\pm$0.02 \\
\hline 

\end{tabular}

\label{alphas}
\end{table*}

\subsubsection{Radio sizes and position angles of the major axes}
\label{morph}

\begin{figure*}
 \centering
 \includegraphics[width=18cm]{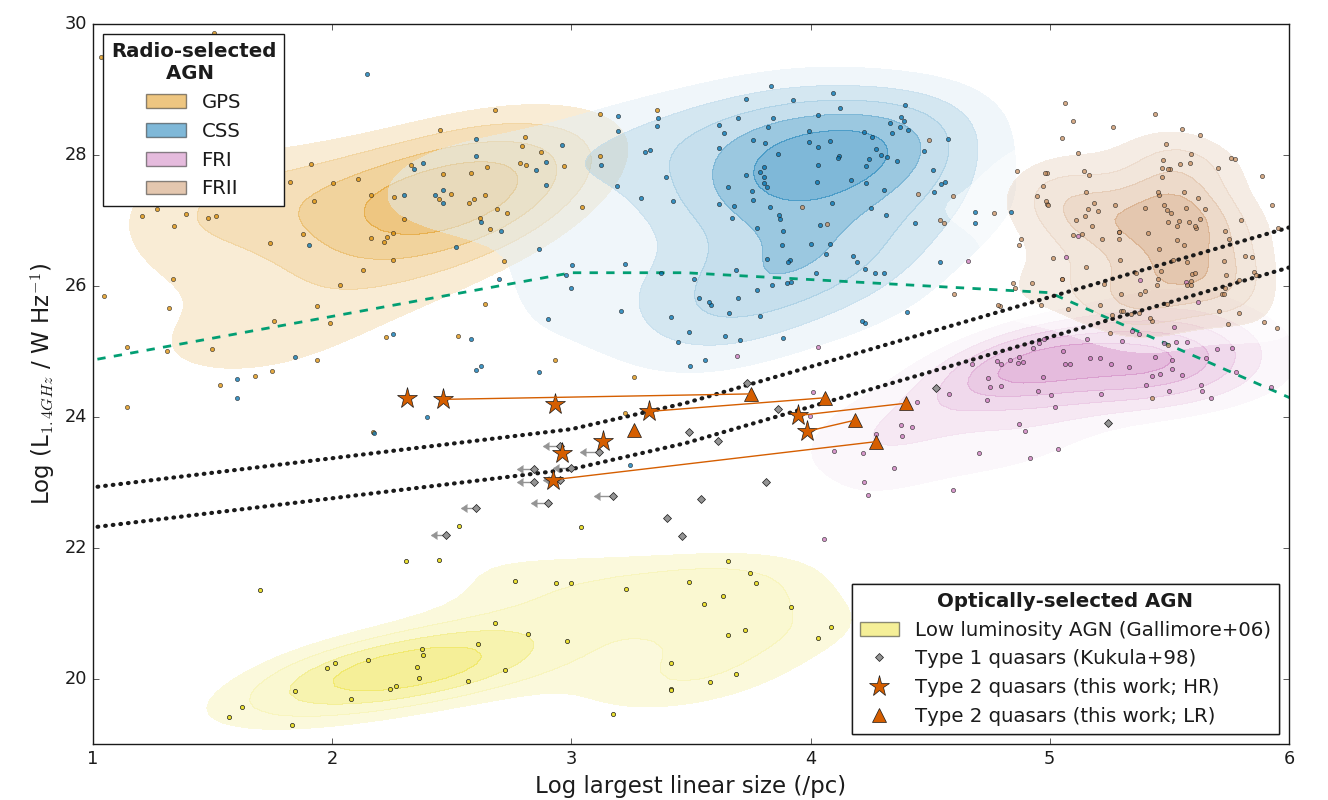}
 \caption{Radio luminosity versus largest linear size for our primary sample. The data for radio-selected AGN are from an extensive sample compiled by \citet{An+12}, shown by points and density clouds of the appropriate colour. Seyfert and LINER galaxies \citep[`low luminosity AGN';][]{Gallimore+06} and type 1 quasars \citep{Kukula+98} are also shown as representative optically-selected AGN for comparison. For our data, we show the sizes and total luminosity from both the high-resolution images (HR; stars) and low-resolution images (LR; triangles). Where both are shown for the same source they are connected by a red line. The two black dotted lines approximately separate the area where laminar jet flows are stable (above the lines) and or unstable and turbulent (below) and the green dashed line shows a possible evolutionary track \citep[see Section~\ref{discussion};][]{An+12}. Our quasars and the type 1 quasars share properties with the lowest luminosity compact radio galaxies (CSS; GPS) and low luminosity small FRI radio galaxies.
  }
  \label{powervsize}
 \end{figure*}

Here we provide two quantitative measures of the large scale radio morphology for each source: (1) the largest linear size (LLS) and (2) the position angle of the major axis. For the size measurements, our method is motivated by comparing to published radio sizes for other samples (see Section~\ref{jets}). Therefore, for both the high and low-resolution showcase images (Fig.~\ref{radio_collage_dat}) the largest linear size is calculated as the distance between the peak emission of the two farthest morphological features \citep[see e.g.][]{Kunert-Bajraszewska+10}. For the features well described by Gaussian components the peak of the Gaussian fits are used and for region components the brightest pixel within the region is used to motivate the peak position. These peak positions are shown as magenta points in Fig.~\ref{radio_collage_dat}. For the components with only one observed morphological feature in our highest resolution radio images (J1010$+$0612, J1338$+$1503 and J1356$+$1026) we used {\sc casa}'s {\sc imfit} function to calculate the deconvolved size. The sizes used for these components are the major axis size from these fits which are 115$\pm$7.8~marcsec ($\sim$0.2~kpc), 595$\pm$41~marcsec ($\sim$2~kpc) and 133$\pm$34~marcsec ($\sim$0.3~kpc) respectively. We plot all these sizes in Fig.~\ref{powervsize}.


To obtain the position angle of the major axis of the radio emission, we defined an axis by a line connecting the two peaks used to calculate the largest linear sizes. To check for consistency with the method used to calculate the position angle of the ionized gas emission (see Section~\ref{line_fitting}), and to estimate the reliability of these position angles, we fit two dimensional Gaussians to the showcase image (Fig.~\ref{radio_collage_dat}) of each target after first applying a Gaussian filter with $\sigma=0.5$~arcsec\footnote{For J1100$+$0846 we used the e-MERLIN image for this and only burred by a $\sigma=0.2$~arcsec Gaussian filter so as to not completely remove the effect of HR:B.}. We then used the difference between the fit PA and that from connecting the two farthest peaks as an error on the position angle. These errors are between $\sim$0.2 and 9$^\circ$ (in all cases this was larger than the formal error on the fit). In most cases, the high-resolution radio images were used to identify the positional angle. The two exceptions are J1338$+$1503 (where we have no high-resolution image) and J1356$+$1026 where an extended feature is only visible in the low-resolution image. In the two cases where no extended radio features were identified in any of our radio images (J1010$+$0612 and J1338$+$1503), the two-dimensional fit in {\sc casa} was used to identify the deconvolved PA and its error. These radio major axes are indicated by the blue dashed lines in Fig.~\ref{radio_collage_dat}.


%

\subsection{Ionized gas maps and analyses}
\label{line_fitting}

Here we describe the steps taken to analyse the ionized gas morphologies and kinematics using our optical IFS data from GMOS and VIMOS (see Section~\ref{IFS}), and how we align these data to our radio maps. We trace the ionized gas kinematics using the [O~{\sc iii}]$\lambda 5007$ emission-line profile and follow the procedures described in detail in \citet{Harrison14,Harrison15}, with brief details given here. 

To map the dominant gas kinematics across the galaxies and compare to the radio morphologies, following \citet{Harrison14,Harrison15}, we use the following non-parametric definitions to characterise the overall [O~{\sc iii}] emission-line profiles:

\begin{enumerate}
\item{The peak signal-to-noise ratio (S/N), which is the S/N of the emission-line profile at the peak flux density. This allows us to identify the spatial distribution of the emission-line gas, including low surface-brightness features.}

\item{The `median velocity' ($v_{50}$), which is the velocity at 50~per cent of the cumulative flux. This allows the `bulk' ionized gas velocities to be traced.}

\item{The line width, $W_{80}$, which is the velocity width that contains 80~per cent of the overall emission-line flux. This characterises the overall width of the emission-line, irrespective of the underlying profile shape. For comparison to other work (Section~\ref{discussion}) we also calculate $W_{90}$, which contains 90~per cent of the overall emission-line flux.}

\item{The asymmetry value \citep[$A$; see ][]{Liu13}, which is defined as: \begin{equation} A \equiv \frac{(v_{90}-v_{50})-(v_{50}-v_{10})}{W_{80}},\end{equation} where $v_{10}$ and $v_{90}$, are the velocities at 10~per cent, and 90~per cent of the cumulative flux, respectively. A very negative (positive) value of $A$ means that the emission-line profile has a strong blue (red) wing. }
\end{enumerate}

\begin{figure*}
 \centering
  \includegraphics[width=18cm]{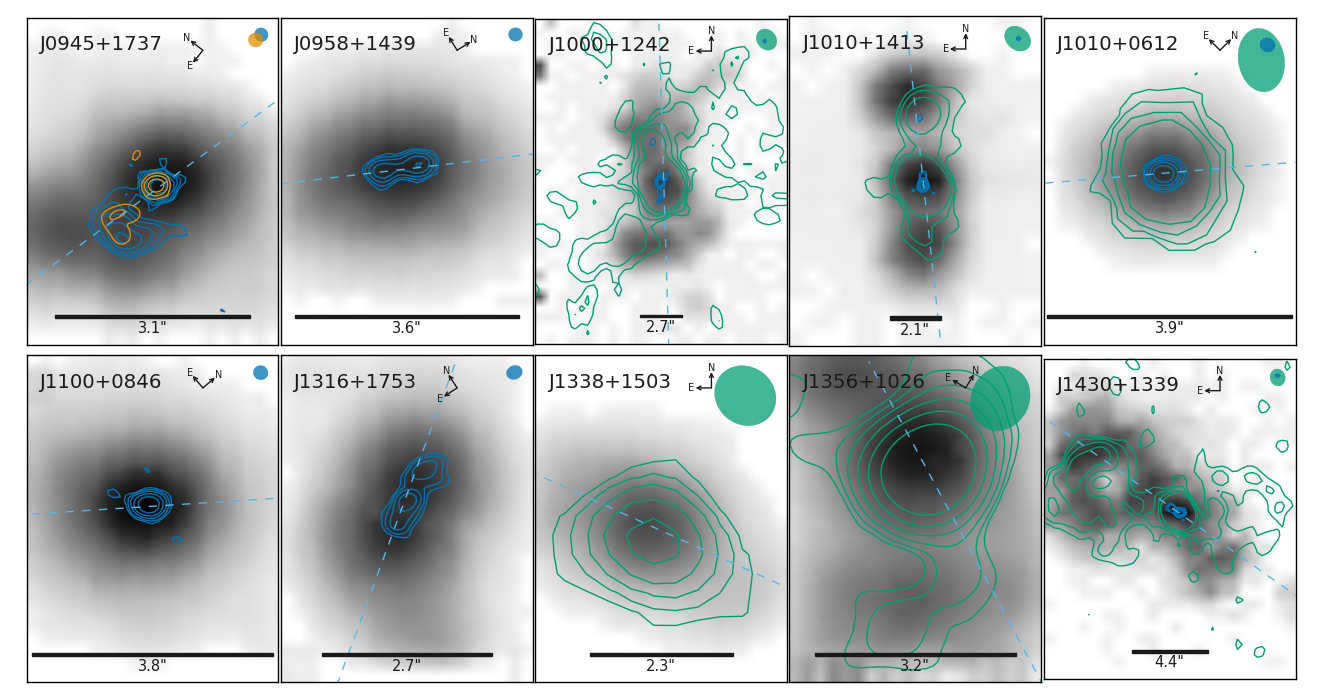}
 \caption{The distribution of [O~{\sc iii}] emission (S/N maps) with contours overlaid from the radio images shown in Fig.~\ref{radio_collage_dat}, colour coded as in that figure. The beam for each radio image is shown as an appropriately coloured ellipse in the top right corner. The dashed line marks the major axis of the [O~{\sc iii}] emission (see Section~\ref{line_fitting}). The scale bar in each panel represents 7~kpc. We observe a close connection between the radio and ionized gas morphologies.}

\label{collage_dat} 
\end{figure*}

To minimize the effect of noise on the broad wings of the emission-lines even in regions of low S/N, we fit the [O~{\sc iii}]4959,5007 emission-line profile with multiple Gaussian components, correcting for the instrumental dispersion, following the methods described in \citet{Harrison14,Harrison15}. We produce maps of each of the parameters described above by fitting the emission-line profiles in $\sim$0.6~arcsec spatial regions (i.e., comparable to the seeing of the observations). The (S/N) maps are shown in Fig.~\ref{collage_dat} and the other maps are presented in Section~\ref{OIII_conection}.

The systemic redshifts quoted in Table~\ref{targets} are derived using the $v_{50}$ values of the [O~{\sc iii}] emission-line profiles extracted from a 3$\times$3~arcsec aperture centred on the quasar's SDSS position in the GMOS data cubes. This corresponds roughly to the velocity of the narrow component, which is often attributed to galaxy kinematics \citep[e.g.][]{2005ApJ...627..721G,2013ApJ...768...75R} or in the case of multiple peaks, lies roughly at the central velocity which, assuming the peaks are dominated by either rotation or symmetric outflows \citep[e.g.][]{Holt2008}, should give a good estimate of the systemic velocity. We note that the values used here vary from the quoted SDSS redshifts by a maximum of $\sim$100\,km\,s$^{-1}$. 

\begin{figure}
 \centering
 \includegraphics[width=\hsize]{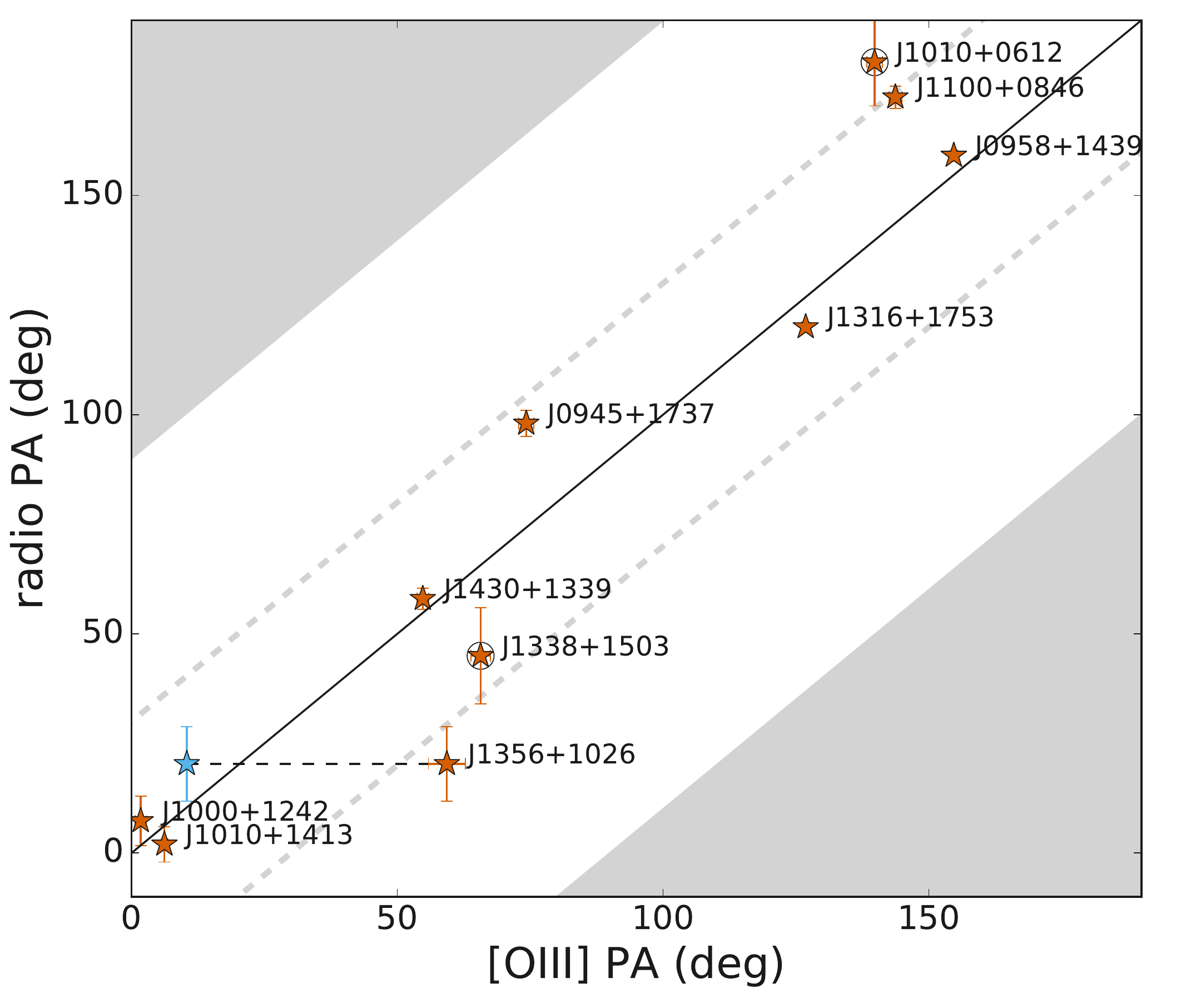}
 \caption{Comparison of the position angle of the major axis of the [O~{\sc iii}] gas and the radio emission for the galaxies in the primary sample (red stars; black circles surround the two soures where only one radio component is observed and the radio PA is from fitting in {\sc casa}). The black line marks x=y, with $\pm30^\circ$ separation marked by grey dashed lines. The areas where the two axes would be separated by $\geq90^\circ$ is shaded out. For J1356$+$1026 a light-blue star, connected to the main point by a black dashed line, marks the position angle of the [O~{\sc iii}] emission if the outflowing bubble is used \citep[which is not covered by our IFS observations;][]{Green12}. There is a close relationship between the spatial distribution of ionized gas and radio emission. }
  \label{PA}
 \end{figure}

To compare the morphology of the ionized gas quantitatively to our radio images, we measure a position angle from our S/N maps. We define the major axis of the [O~{\sc iii}] emission by fitting a single Gaussian to the S/N map for each galaxy. This method is slightly biased to passing through the brightest features in the [O~{\sc iii}]; however, based on visual inspection provides a sufficient measurement of the position angle for our broad comparison to the distribution of radio emission presented in Section~\ref{OIII_conection}. The main exception is J1356$+$1026 for which the position angle we measure for the ionized gas is determined primarily by a bright region to the north east of the core with no radio counterpart observed. However, if we define the position angle using the location of the base of the bubble identified by \citet{Green12} (see Section~\ref{OIII_conection}), the radio and [O~{\sc iii}] emission are well aligned, with a PA separation of $\sim$10\,degrees. These positional angles are shown on the [O~{\sc iii}] S/N maps in Fig.~\ref{collage_dat} and are plotted against the radio PAs in Fig.~\ref{PA}. 


%


We aligned the IFS data to the SDSS astrometry by creating pseudo-broad band images from the IFS cubes using the common wavelength coverage with SDSS \emph{r} band\footnote{for J1338$+$1503 the \emph{g} band filter was used because the \emph{r} was too effected by bad pixels near the edge of the IFS wavelength range.}. We then anchored the astrometric information of the IFS peak pixel location to the associated position from the SDSS image peak, after blurring each with a 0.2~arcsec Gaussian filter to minimize the impact of bad pixels. We used the brightest morphological components in our high-resolution VLA images (HR:A) to confirm that the SDSS and VLA astrometry were consistent\footnote{J0958$+$1439 was excluded from this since it seems likely, due to their similar brightness and steep spectral indices, that both of the radio components in this source are lobes around a central, core below our detection threshold.}. We found the SDSS positions were scattered around the peak VLA emission with a median offset of 0.13~arcsec, corresponding to 0.24~kpc at a representative redshift of $z$=0.1. This is sufficient for the comparison we make between the radio morphologies and our $\sim$\,0.6--0.7\,arcsec resolution ionized gas kinematics in this work (Section~\ref{OIII_conection}).


\section{Results and discussion}
\label{results}

In the previous section we presented our 1--7\,GHz radio imaging and optical integral field spectroscopy for a sample of ten $z$$<$$0.2$ `radio-quiet' quasars, that were known to host $\gtrsim$kpc-scale ionized outflows based on our previous work \citep[Fig.~\ref{collage_dat};][]{Harrison14,Harrison15}. The aim of this section is to establish the origin of the radio emission (Section~\ref{origin_of_radio} and Section~\ref{sec:origin}), to explore the relationship between the radio and the ionized gas (Section~\ref{OIII_conection}) and to discuss the implication of our results for understanding the radio emission and feedback in the context of the overall AGN population (Section~\ref{discussion}). 


\subsection{Properties of the observed radio emission}
\label{origin_of_radio}

We observe radio structures with a range of morphologies from compact features with spatial extents of $\sim$1\,kpc (e.g., see the high-resolution image of J0958$+$1439 in Fig.~\ref{radio_collage_dat}) to diffuse lobes extending over $\sim$25\,kpc (e.g., see the low-resolution radio image of J1000$+$1242 in Fig.~\ref{radio_collage_dat}). In particular five of our targets show distinctly jet like radio morphologies (J0945$+$1737, J0958$+$1439, J1000$+$1242, J1010$+$1413 and J1316$+$1753) with three more showing more irregular radio features \citep[J1100+0846, J1356+1026 and J1430+1339; see Section~\ref{sec:classification} and ][]{Harrison15}. Specifically, this means that of the nine quasars in the sample consistent with being radio excess in Section~\ref{characterising} (i.e.~all except J1338$+$1503; see Fig.~\ref{radio_loudness} and Table~\ref{SED_vals}) 90~per cent show spatially resolved radio structures with linear sizes on $\sim$1--25\,kpc scales (see Fig.~\ref{radio_collage_dat} and Fig.~\ref{powervsize}).

To estimate the significance of the features that we have identified in our high-resolution VLA and e-MERLIN data in terms of their contribution to the total radio luminosity at 1.5~GHz, we compare the radio emission from these morphologically-distinct features to the total radio emission (extracted from FIRST but consistent with our observations, see Section~\ref{radio_SED}; Table~\ref{targets}). For the radio excess sources, we find that the total combined fluxes of the high-resolution components, including the central nuclear components, (i.e., HR:Total in Table~\ref{alphas}) contain $\sim$60--90~per cent of the total radio flux. The exception is J1430$+$1339 for which the high-resolution components only make up $\sim$22~per cent of the total flux with $\sim$50~per cent of the FIRST flux located in the diffuse low-resolution lobes/bubbles. Below, we discuss how the radio emission that is resolved out on these $\sim$0.25~arcsec scales may be attributed to star formation.

Importantly, as can be seen by eye in Fig.~\ref{sdss} and quantitatively using the $\Theta_{FIRST}$ parameter (see Table~\ref{targets}), only J1430$+$1339 is definitively extended and two other sources (J0945$+$1737 and J1000$+$1242) are tentatively extended based upon their $\sim$5\,arcsec resolution FIRST data. This is supported by the `FIRST Classifier' \citep{Alhassan18}, which automatically identifies FIRST sources as compact or not, and determines that all of our sources are compact except for J1430$+$1339. This cautions against only relying on low-resolution radio data to identify low power/compact radio structures not associated with star formation in such systems \citep[e.g., see][]{Kimball+11,Le+17}.

\subsection{Origin of the radio emission}
\label{sec:origin}

\subsubsection{Star formation}
\label{sec:star_formation}
All of our targets are classified as being `radio-quiet' based on standard criteria \citep[e.g.,][see Fig.~\ref{radio_loudness}]{Xu+99}. Furthermore, based on many standard criteria, our sources would {\em not} be classified as `radio AGN' \citep[e.g.,][ Section~\ref{characterising}]{Best&Heckman}. The radio emission in such sources is often attributed to being dominated by star formation processes \citep[e.g.,][]{Best&Heckman,Condon13}. However, through unresolved \emph{UV}--to--\emph{FIR} SED fitting we found that nine of our ten type~2 quasars have more radio emission than can be explained from star formation alone (Section~\ref{characterising}). For these nine targets, star formation estimates from SED fitting imply that only $\sim$3--11~per cent of the observed radio emission at 1.4~GHz are produced by star formation (see Table~\ref{SED_vals}). 

A comparison of the total 1.4\,GHz flux density in our high-resolution images to the total flux obtained from FIRST reveals that $\sim$10--40~per cent of the total radio flux is resolved out across the sample. In all cases, the amount of flux resolved out in the high-resolution images is greater than the 1.4\,GHz flux predicted from our calculated star-formation rates using the \emph{FIR}--radio correlation \citep[see Table~\ref{SED_vals};][]{Bell+03}. This means that the radio emission from star formation can be fully accounted for with a diffuse component not identified in our high-resolution images. Although convincing, we note that these arguments are based upon SED fitting results which are subject to some systematic uncertainties (see Section~\ref{characterising}). 

Another piece of evidence that star-formation does not dominate the radio emission in the nine radio excess targets is their complex radio morphologies \citep[see Fig.~\ref{radio_collage_dat}; e.g.][]{Colbert96}. However, it is plausible that star formation could contribute to the central/nuclear emission we see in our high-resolution images. We assess this possibility independently of our SED fitting results. We initially measured the nuclear (HR:A) radio sizes from two-dimensional beam-deconvolved Gaussian fits on the high-resolution images shown in Fig.~\ref{radio_collage_dat} using {\sc casa}. We obtain major axis sizes of $\sim$100--200\,marcsec, with errors at least 4 times lower than these values. If we then {\em assume} that all of the radio emission observed is due to star formation, following \citet{Bell+03} and \citet{Kennicutt12}; corrected to a Chabrier IMF \citep{Chabrier03} the inferred SFR surface densities are consequently $ \log \Sigma\approx 2$--4[M$_{\odot}$/yr/kpc$^2$]. These values straddle the physical cut-off set by the Eddington limit from radiation pressure on dust grains (i.e., $\log \Sigma \approx3.5$), with five of the nine sources lying above the Eddington limit \citep{Murray05,Thompson05,Hopkins10}. These results strengthen our SED-based arguments that the radio structures observed in our high-resolution images, including the nuclear components, are not dominated by star formation processes. 

In summary, for all but one of our ten targets we have strong evidence that only $\sim$3--11~per cent of the total flux can be attributed to star formation. All of the radio structures that we see in the high-resolution images appear to be dominated by other processes associated with the AGN.

\subsubsection{Quasar winds}
\label{winds_section}

Another largely discussed source of the radio emission in radio-quiet quasars is radiatively-driven accretion disc winds which result in synchrotron emitting shocks through the inter-stellar medium \citep[e.g.][]{Jiang10,Zakamska14,Nims+15,Zakamska16,Hwang18}. Currently there are only rough predictions for this scenario and these are only for spatially-integrated radio properties (i.e., not spatially-resolved). One prediction we can use is that the simple energy conserving outflow model presented by \citet{FaucherGiguere12}, when launched by a quasar with $L_{\rm AGN}$~$\approx$10$^{45}$\,erg\,s$^{-1}$, could plausibly produce radio luminosities consistent with our targets (i.e., $L_{\rm 1.4GHz}$~$\approx$10$^{23}$--10$^{24}$\,W\,Hz$^{-1}$) when it interacts with the interstellar medium \citep{Nims+15}. These modelled outflows can reach velocities of $\sim$1000~km~s$^{-1}$, consistent with those seen in our IFS data (see Section~\ref{OIII_conection}). The predicted steep-spectral index from this model ($\alpha$$\approx$$-1$), is also broadly consistent with many of the radio features seen in our observations \citep[see Fig.~\ref{alpha_vs_dist} and Table~\ref{alphas};][]{Jiang10}. However there are a large number of assumptions needed to obtain this conclusion and, importantly, we can now use our high-resolution radio data to further investigate quasar winds as the producer of the radio emission in these quasars.

An outflow driven by a quasar wind may produce loosely collimated radio structures on large scales due to the galactic disc collimating the outflow \citep{Alexandroff+16}. However, the shocked wind scenario described above does not seem sufficient to explain the highly collimated radio structures seen in our high-resolution radio images (e.g., particularly see J1000$+$1242 and J1010$+$1413 in Fig.~\ref{radio_collage_dat}). Furthermore, for J0958$+$1439, the radio structure appears to be directed {\em into} the disc (based on the SDSS morphology and [O~{\sc iii}] kinematics; see Fig.~\ref{sdss} and Fig.~\ref{kinematics}) which also disfavours the wind scenario, but be consistent with randomly oriented radio jets \citep[e.g.,][]{Gallimore+06,Kharb+06}. Unfortunately, it is challenging to identify the galaxy major axes based on the available imaging for most of our sources (see Fig.~\ref{sdss}) and our optical IFS data of the targets are not deep enough to model the orientation of the stellar discs \citep{Kang18}. Another alternative to asses the relative orientations would be to identify a molecular galactic disc in these systems using resolved CO observations \citep[Thomson+in prep;][]{Sun2014}. 

\subsubsection{Jets}
\label{jets}

Given the ubiquity of jets in radio-loud AGN it is reasonable to assume that low luminosity jets can, at least, contribute to the radio emission observed in AGN with lower radio powers. Indeed, radio jets can be identified in `radio-quiet' Seyfert galaxies when using sufficiently deep and high-resolution radio observations \citep{Gallimore+06,Baldi18}.

Our sample of `radio-quiet' quasars (see Fig.~\ref{radio_loudness}) have many properties in common with jetted radio-loud AGN. Specifically, the radio morphologies of our targets as seen in our high-resolution images (Fig.~\ref{radio_collage_dat}), in general, look very similar to jetted compact radio galaxies, with a combination of hot spots, jets and cores \citep[e.g.,][]{Kimball+11,Baldi18}. The jet interpretation is particularly strong for J1000$+$1242 and J1010$+$1413 due to the presence of compact, flat spectrum components \citep[i.e., $\alpha\gtrsim-$0.6; likely to be hot spots; see e.g.][]{Meisenheimer89,Carilli91} inside the more diffuse steep spectrum lobes which are apparent in our low-resolution images (see Fig.~\ref{radio_collage_dat}). For the more compact jet-like structures that we see (e.g. J0958$+$1439 and J1316$+$1753) we would require higher spatial resolution images to separate out possible hot spots from steep spectrum lobes (see Table~\ref{alphas}). 

To quantify our comparison to the traditional radio AGN population, we investigate the radio size (LLS; see Section~\ref{morph}) versus radio luminosity plane for our sources and a literature compilation of radio selected AGN from \citet{An+12} in Fig.~\ref{powervsize}. In terms of radio luminosity, our targets are consistent with the lowest luminosity radio-identified AGN samples \citep[e.g.,][]{Fanti87,Kunert-Bajraszewska+10} and fill in the gap between these `radio-loud' AGN and Low-Luminosity AGN \citep[e.g.][]{Gallimore+06}. Based on our low-resolution images, where we can see $\sim$6--20\,kpc radio structures, four of our targets overlap with Fanaroff-Riley class I \citep[FRI;][]{Fanaroff74} galaxies in the luminosity--size plane. However, the morphologies that we observe in our targets are not clearly consistent with this class of objects, which are more dominated by `lossy' jets and have relatively weaker hot spots \citep[e.g., as seen in 3C 31;][]{Laing08}. We discuss possible reasons in Section~\ref{discussion}. 

It can be seen in Fig.~\ref{powervsize} that most of our sources have radio sizes spanning those seen in compact steep spectrum (CSS) radio galaxies \citep[i.e., $\sim$1--25\,kpc;][]{ODea98}. In the most compact case of J1010$+$0612, we see no features beyond the nuclear component and the deconvolved radio size is $\sim$200\,pc, such that it is more consistent with those seen in Gigahertz Peak Spectrum (GPS) objects \citep{ODea98}. Interestingly, there is an observed relationship between radio sizes and the frequency of peak emission in the radio SEDs of compact radio galaxies \citep{Orienti+14}. Within our limited ability to identify a turnover in the radio SEDs and to constrain the turnover frequency, our targets are consistent with this relation, with three or four of our targets in particular showing a turnover in the radio SEDs somewhere between FIRST (1.4~GHz) and TGSS (150~MHz; see SEDs in Fig.~\ref{SED} and the supplementary information)\footnote{These are J1000$+$1242, J1100$+$0846 and J1356$+$1026, and possibly J1010$+$0612}. Further multi-frequency radio observations are required to accurately identify the turnover frequencies in our targets.

In a few cases, we see that the brightest nuclear radio component has a moderately flat spectral index (i.e., $\alpha >-0.6$), which may indicate a contribution from radio emission associated directly with an AGN `core' / accretion disc \citep{Padovani16}. Although we do not see strong evidence of flat spectrum AGN cores across the full sample, with most sources showing steep spectral indices in their nuclear regions, there are several possible explanations. For example, the radio core could have recently turned off which would cause its spectral slope to steepen and simultaneously could explain their low radio luminosities \citep{Kunert-Bajraszewska+10}. Alternatively, multiple episodes of jet activity would produce a similar effect with the unresolved, younger jets/lobes outshining the core \citep[e.g.][]{Kharb+06,Gallimore+06,Orienti16}. Higher resolution images, particularly at higher frequencies, where the relative contribution from a flat spectrum core would be higher, are required for a thorough search for radio cores in our targets \citep[see e.g.][]{Middelberg04}.

\subsubsection{Final classification of radio features}
\label{sec:classification}

Our final classifications of the radio structures that we have observed are given in the `interpretation' column of Table~\ref{alphas}. These classifications are based on the morphology, spectral index and distance of the features from the optical centre (Fig.~\ref{alpha_vs_dist}). 

We have presented multiple pieces of evidence that support a jet origin for the majority of the non-nuclear morphological radio features we observe in our targets. However, for the nuclear, central components that have steep spectral indices (i.e., $\alpha<-0.6$), it is plausible that some fraction of the radio emission could be due to radiative winds that have shocked the interstellar medium (Section~\ref{winds_section}). Only in J1338$+$1503 can we not rule out that star-formation dominates the radio emission.

In the high-resolution components the name `nuclear' was applied to the component closest to nucleus (based on the SDSS position), which in every case was also the brightest radio component (HR:A). We further split the nuclear components by either having a core contribution or being jet/lobe/wind dominated based on whether its spectral index was steep ($\alpha<-0.6$) or flat ($\alpha>-0.6$), respectively (see Fig.~\ref{alpha_vs_dist})\footnote{For J0958$+$1439 whose two roughly equal brightness components are approximately equidistant to the optical centre (0.6 vs 0.8~kpc), both are labelled jet / lobe.}. The non-nuclear high-resolution components are labelled either as jet/lobe or as hot spot depending on if they are steep or flat \citep[see e.g.][]{Dallacasa13}. 

There are a few exceptions to these clean divisions of the high-resolution components, which we label as `unclear' in Table~\ref{alphas}. In J0945$+$1737, HR:C and HR:D are low signal-to-noise features, which may not be truly individual components. Furthermore, HR:B in J1100$+$0846 which only appears in the e-MERLIN image, is either an extremely high significance artefact or a variable component which is below the detection limit at the epoch of the high resolution VLA observations. Assuming that it is real and non-variable would require a nonphysical spectral index of $\lesssim$$-$$4$ using the 5$\sigma$ upper limits given in Table~\ref{alphas}. The artefact explanation is supported by HR:B containing $\sim$30~per cent of the peak flux, which is comparable to the highest peak in the synthesised beam; however in this case we would expect a symmetric feature on the other side of the core (that we do not see). Assuming a conservative, yet not un-physical, spectral index of $\lesssim-2$ \citep{Harwood17} for HR:B over a $\sim$1~year period, a factor of ten variability at 5.2~GHz would be required for it to be un-detected in our VLA image (see Table~\ref{alphas}). Similar scale (1-2 orders of magnitude) variability has been seen on month-year time-scales in radio-quiet quasars \citep{Barvainis05}. Further multi-epoch observations would be needed to confirm this interpretation.

For the low-resolution radio components, they are generally either labelled as composites, when they are composed of multiple observed high-resolution components, or lobes. Additionally, LR:A in J1338$+$1503 is classified as probably star formation dominated due to the lack of high-resolution data for this target and it not being classified as radio excess. The final exception to the low-resolution classifications is LR:B in J1356$+$1026. Its lack of any high-resolution counterpart makes its identification as a jet or a lobe more tenuous.

\subsection{Connection between ionized gas and radio}
\label{OIII_conection}

 \begin{figure*}
 \centering
 \includegraphics[width=17cm]{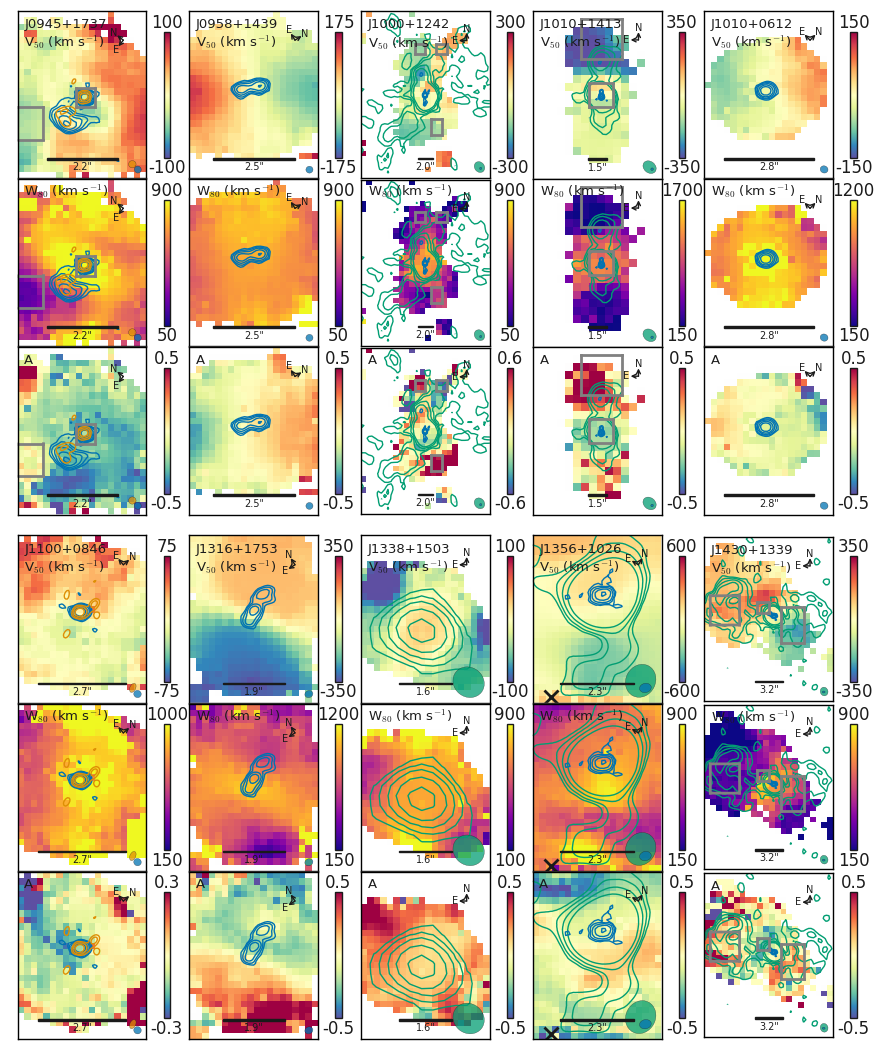}
 \caption{ For each of the targets in our primary sample, we show maps of three non-parametric emission-line properties measured from the [O~{\sc iii}] profile in our optical IFS data (see Section~\ref{line_fitting}). These are given in columns for each target starting with the `median velocity' ($v_{50}$), then the line width ($W_{80}$) and the asymmetry ($A$), with each having their respective scale bar shown to the right. In each case, the relevant radio contours are overlaid with the low-resolution VLA in green, the high-resolution VLA in blue and e-MERLIN in orange; in each case the beam(s) are shown in the lower right corner. The scale bar in each image represents 5~kpc. Relevant features / regions discussed in Section~\ref{OIII_conection} are shown as either grey boxes or black crosses; [O~{\sc iii}] emission-line profiles extracted from each of these boxes are shown in the online supplementary material.
 }
  \label{kinematics}
 \end{figure*}
%

We have previously identified $\gtrsim$kpc scale ionized gas outflows in our sample of type~2 quasars \citep{Harrison14,Harrison15}. With our new radio images, we are now in a position to compare the radio morphology with the morphological and kinematic structures of the ionized gas that we can obtain from our IFS data of each of these targets (Section~\ref{line_fitting}). 


\subsubsection{Morphological alignment of radio and ionized gas}
\label{OIII_align}

In Fig.~\ref{collage_dat}, we compare the spatial distribution of the ionized gas, as traced by [O~{\sc iii}], to the distribution of radio emission in each of our ten targets. Specifically for J0945$+$1737, J1000$+$1242, J1010$+$1413 and J1316$+$1753, where we see extended, distinct ionized gas structures, we see [O~{\sc iii}] bright regions in front of, or co-spatial with, the hot spots and jet-like features we identified in Section~\ref{radio_SED}. Additionally in J1430$+$1339 we see co-incident bubbles of radio emission and ionized gas \citep[discussed in detail in][]{Harrison15,Lansbury18}. Finally, in J1356$+$1026, we see a radio structure that is difficult to classify (Section~\ref{sec:classification}), but is a possible radio jet/lobe. It is located at the top of a $\sim$12\,kpc [O~{\sc iii}] bright region extending to the south, not covered by our IFS data, but clearly seen in {\em HST} imaging (see supplementary material) and confirmed to be an outflowing bubble by \citet{Green12}. 

The alignment between the ionized gas and radio emission in all ten of our targets is quantified in Fig.~\ref{PA} by comparing the position angle of the semi-major axis of the radio data (blue dashed lines in Fig.~\ref{radio_collage_dat}) and ionized gas (blue dashed line in Fig.~\ref{collage_dat}). We find that nine targets have alignments within 30\,degrees. The exception is J1010$+$0612, which shows no distinct morphological features on $\sim$\,kpc scales in either our radio or [O~{\sc iii}] images.

Similar alignments between radio emission and ionized gas have been seen by other studies of `radio-quiet' quasar and Seyfert populations. These are generally interpreted as the radio jet interacting with the ISM, causing outflows, bow-shocks and sometimes deflecting the jet \cite[e.g.,][]{Ulvestad83,Ferruit99,Whittle04,Leipski06}.

\subsubsection{Connection between jets and ionized gas kinematics}
\label{OIII_kinematics_conection}

In Fig.~\ref{kinematics}, we overlay our radio images on top of the kinematics maps from our IFS data (described in Section~\ref{line_fitting}). We provide further visualisations of the [O~{\sc iii}] emission-line profiles at the locations of the radio structures in the supplementary online material. We defer a detailed kinematic analyses of the ionized gas in our targets, and a quantitative comparison to e.g., jet power to future work \citep[also see][]{Harrison15}. Here we provide a first overview of the relationship between the large-scale kinematic properties of the warm ($\sim$10$^{4}$\,K) ionized gas and the radio features we identified in Section \ref{radio_SED}.

It can be seen in Fig.~\ref{kinematics} that the ionized gas shows distinct kinematics at the location of the spatially-extended jet/lobe structures that we have identified in our sources. For J1430$+$1339, we have already presented (in earlier work) the presence of a broad, high-velocity ionized gas component (W$_{80}$$\approx$900\,km\,s$^{-1}$ and $v_{p}$$\approx$600\,km\,s$^{-1}$; marked by a small central grey box in Fig.~\ref{kinematics}) co-spatial with the HR:B radio jet/lobe structure \citep{Harrison15}. In addition, the $\sim$20\,kpc scale bubbles observed in both the ionized gas and radio emission (see Fig.~\ref{collage_dat}; also marked by grey boxes in Fig.~\ref{kinematics}) have narrow emission-line profiles but offset velocities possibly indicative of outflows ( $\approx \pm$150~km s$^{-1}$ respectively). Coronal line measurements in these regions by \citet{Villar-Martin18} confirm this velocity offset and suggest that the north-east bubble may contain ionization level dependent kinematic substructure.

The potential jets we observe in J0945$+$1737 and J1010$+$1413 (HR:B and HR:C respectively), terminate at brightened blue-shifted [O~{\sc iii}] clouds (with $v_{p}=-27$ and $v_{p}=-316$~km s$^{-1}$, respectively; see grey boxes in Fig.~\ref{kinematics}). This is evidence of jets hitting a cloud of gas, both pushing the gas away and deflecting the jet \citep[see e.g.][]{Leipski06}. For J1010$+$1413 this supports the interpretation that the ionized gas region in the north is part of an outflow rather than being passively illuminated by the AGN \citep[see also][]{Sun17}. In J1316$+$1753, we see strong double peaked [O~{\sc iii}] emission, offset in velocity by $\sim$400\,km\,s$^{-1}$, with the blue and red shifted gas being brightest at the termination of each jet (also see additional figures in the supplementary material). J0958$+$1439 shows a similar kinematic line splitting structure and co-spatial jets/lobes. Such observations indicate possible jet-driven outflows similar to that seen in \citet{Rosario10}.

\begin{figure}
 \centering
 \includegraphics[width=\hsize]{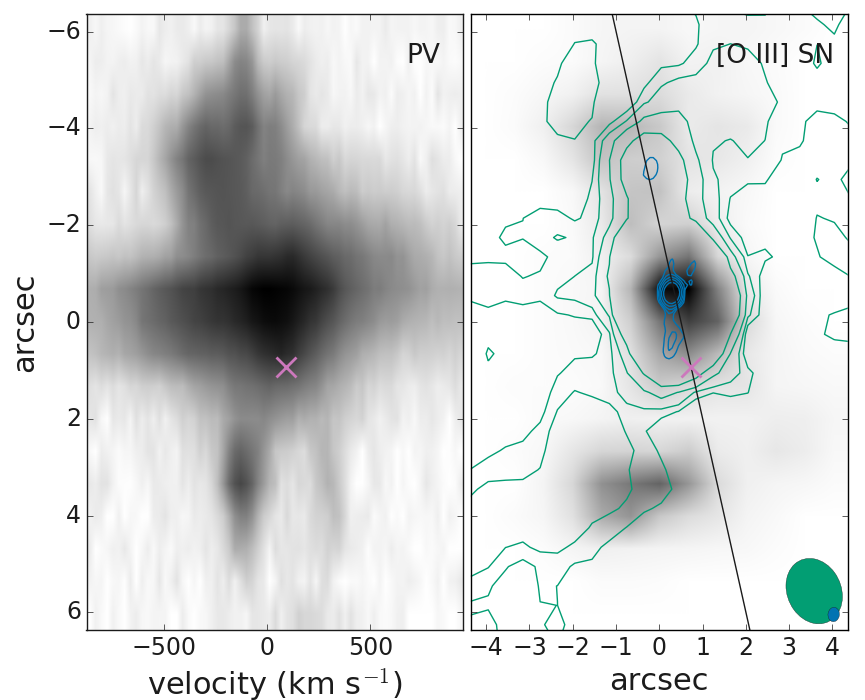}
 \caption{ A position velocity diagram for J1000$+$1242 (left) extracted along a 2~arcsec wide pseudo-long slit from our VIMOS data with the slit position shown as the black line on the [O~{\sc iii}] S/N map, with radio contours from Fig.~\ref{radio_collage_dat} over-plotted (right). We see the signature line splitting of a quasi-spherical outflow starting approximately at the pink `x' in both panels. This bubble seems to begin roughly at the base of a probable radio jet (HR:B). We have also identified a radio structure at the base of a similar outflowing bubble in J1356$+$1026 (identified by \citet{Green12}; see black `x' in Fig.~\ref{kinematics}).}
  \label{J1000_bubble}
 \end{figure}

Striking evidence of outflowing bubbles of gas being launched near the base of likely jets/lobes is observed in both J1000$+$1242 and J1356$+$1026 (HR:B and LR:B, respectively; Fig.~\ref{kinematics}). For J1000$+$1242, this is seen in the kinematic maps by the blue-to-red $\sim$200\,km\,s$^{-1}$ velocity shift in the $v_{50}$ map from east--west (extracted from the two northerly regions in Fig.~\ref{kinematics}) and the large asymmetry values in both the north and south (A$\approx$1). This bubble is characterised by velocity splitting of the [O~{\sc iii}] emission-line seen using a pseudo-slit extracted from our IFS data in Fig.~\ref{J1000_bubble}. We find that the base of this southern bubble corresponds to the location of the southern jet (HR:B). No sign of this bubble was seen in long-slit observations with a similar alignment in \citet{Sun17}, possibly due to the longer exposure time and better spectral resolution of our data. A $\sim$12\,kpc outflowing bubble in J1356$+$1026 was discovered by long-slit observations in \citet{Green12}, with a very similar kinematic structure to the one we see for J1000$+$1242 in Fig.~\ref{J1000_bubble}. For J1356$+$1026, the bubble is beyond the field-of-view covered by our GMOS data cube. However, here we have discovered a radio feature that terminates at the base of the outflowing bubble (LR:B; see the black `x' in Fig.~\ref{kinematics}). Although the origin of this radio structure is ambiguous (see Section~\ref{sec:classification}), this may also be due to a jet that terminates at this location and drives the outflow. 

We note that the spatial extent of the outflows in many of our sources (J1000$+$1242, J1010$+$1413, J1356$+$1026 and J1430$+$1339 in particular) are underestimated, if the outflow size is based solely upon the spatial extent of the broad [O~{\sc iii}] emission-line component \citep[see e.g.,][]{Kang18} and highlights the need for careful analysis when establishing outflow properties \citep[e.g., see][]{Harrison18}.

In summary, we observe a strong relationship between the radio jets/lobes and the ionized gas kinematics in all seven of the targets where we see unambiguous radio structures on 1--25\,kpc scales. 




%
 
\subsection{Radio jets associated with quasar outflows and feedback}
\label{discussion}

 \begin{figure}
 \centering
 \includegraphics[width=\hsize]{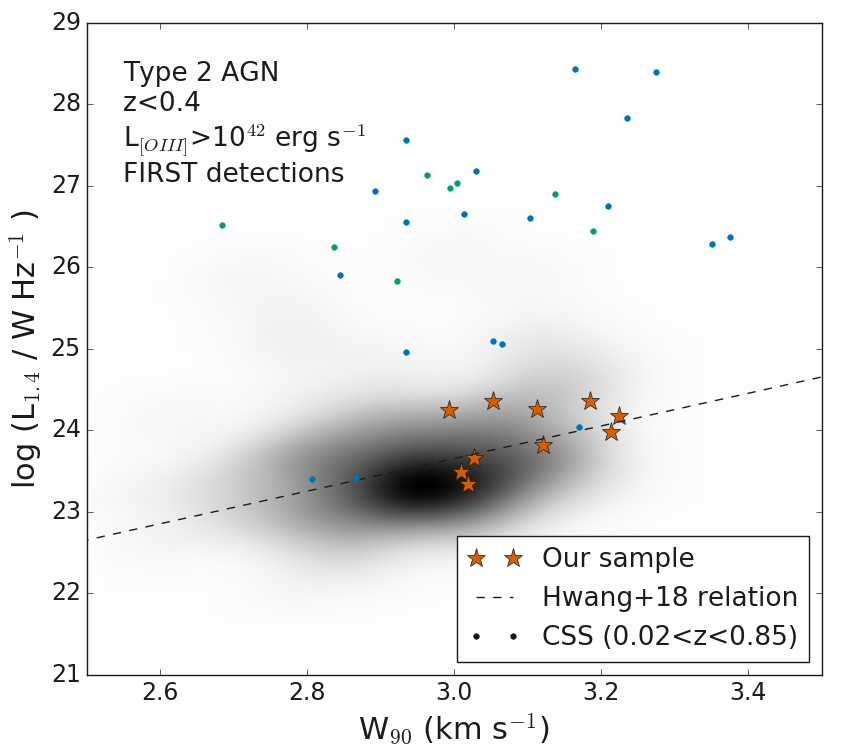}
 \caption{The width of the [O~{\sc iii}] line (W$_{90}$) as a function of radio luminosity. The grey-scale shows the underlying distribution of type 2 AGN \citep[][]{Mullaney13} with $z$$<$$0.4$ and a luminosity cut of $L_{\text{ [OIII]}}>10^{42}$~erg s$^{-1}$ (i.e., the selection criterion of our sample; Fig.~\ref{targets}) that are detected in FIRST. A similar selection criterion was used by \citet{Hwang18}, using the $z$$<$$0.8$ quasars from \citet{Zakamska14}, to produce the relationship shown by the dashed line. Our sample is shown as red stars, where W$_{90}$ is measured from our GMOS data in a 3~arcsec aperture (approximating the SDSS fibre). Circles mark the CSS radio AGN from \citet{Holt2008} (green; using the FWHM of their single Gaussian component fits), and \citet{Gelderman94} (blue). Our sources have more representative radio luminosities compared to CSS radio galaxies; however, we have shown that they also contain compact radio jets. Consequently, radio jets should be explored as a crucial feedback mechanism for all quasars.}
  \label{width_v_lum}
 \end{figure}
%

In this section, we put our work into the context of observational and theoretical studies of other AGN and quasars. In particular, we focus on the properties of the likely radio jets in our sources compared to other samples and theoretical predictions. We explore the implication of our results for understanding the impact of jets on the host galaxies of our targets, and how that relates to the quasar population as a whole. 

Our sources share many properties with the typically more powerful, compact radio galaxies (see Section~\ref{jets}; Fig.~\ref{powervsize}). It has been postulated that the compact radio galaxies (with $\lesssim$10\,kpc scale jets) could evolve into traditional $\approx$100\,kpc double radio galaxies \citep[e.g., see discussion in][and the dashed track in Fig.~\ref{powervsize}]{An+12}. However, not all compact radio sources may be destined to evolve into traditional radio galaxies. Of particular relevance here is the idea that jets can get frustrated/stagnated by the dense interstellar medium \citep[e.g.,][]{vanBreugel+84,ODea91,Bicknell18} and that these jets may consequently become unstable at small sizes \citep[see dotted tracks in Fig.~\ref{powervsize};][]{An+12}. Our targets, along with a sample of type 1 quasars from \citet{Kukula+98} straddle this instability criterion at sizes of only a few kpc (Fig.~\ref{powervsize}). Within the model presented by \citet{An+12}, a characteristic of obstructed jets would be a hot spot and a plume-like diffuse structure beyond the hot spot -- a strikingly good description of what we sometimes see in our targets, in particular for J1010$+$1413 and J1000$+$1242 (Fig.~\ref{radio_collage_dat}). We see further evidence that jets are interacting with the interstellar medium in their host galaxies due to the highly disturbed ionized gas, outflowing bubbles and brightened [O~{\sc iii}] structures observed co-incident with the jets/lobes (Fig.~\ref{collage_dat} and Fig.~\ref{kinematics}; Section~\ref{OIII_conection}). 


It has been observed for several decades that jets interact with their interstellar medium in `radio-loud' samples as well as `low-luminosity' AGN and Seyferts \citep[e.g.][]{vanBreugel+84,Whittle+86,Pedlar89,Capetti96,Steffen97,Ferruit98,Mahony13,Riffel14,Morganti15,Rodrıguez-Ardila17,Nesvadba17,May18,Morganti18}. It has further been noted that compact radio galaxies may host the most extreme ionized gas kinematics because the radio jets are confined in the interstellar medium \citep[e.g.,][]{Holt2008}. In this work, we have provided observational evidence that compact radio jets may be crucial for interacting with the interstellar medium and driving outflows, even in `radio-quiet' quasars, sources where radiatively driven winds are often assumed to be the most important \citep[e.g.,][]{Zakamska14,Hwang18}.

\citet{Hwang18} suggest that the source of the radio emission in radio-quiet quasars (suggested to be winds) is distinct from their radio-loud counterparts (suggested to be jets). This is largely based on a sample of radio-quiet AGN for which the width of the [O~{\sc iii}] and the radio luminosity are roughly correlated, while the jetted CSS sources from \citet{Holt2008} have more radio emission for a given [O~{\sc iii}] width (see Fig.~\ref{width_v_lum}). However, these results are based on low-resolution radio data, which are insensitive to small-scale jets. We find that our targets lie on the relationship seen by `radio-quiet' sources and we have presented several pieces of evidence that our sources contain jets that are interacting with the interstellar medium. Furthermore, the lower power end of the sample of jetted CSS sources studied by \citet{Gelderman94} also lie on the `radio-quiet' relationship in Fig.~\ref{width_v_lum}.

Cutting-edge models show that compact jets interacting with the interstellar medium may be a crucial aspect of `AGN feedback' and possibly the most efficient mechanism for driving powerful outflows \citep[e.g.,][]{Wagner12,Mukherjee16,Bicknell18,Cielo+18}. For example, \citet{Mukherjee18} show that jets can increase the turbulence of the gas within the disc and simultaneously drive larger-scale outflowing bubbles, in qualitative agreement with the observations presented here for some of our targets (e.g., J1000$+$1242; J1356$+$1026; J1430$+$1339; see Fig.~\ref{kinematics} and \ref{J1000_bubble}). In future work, we will use our IFS data and radio imaging to measure the detailed outflow energetics in relation to the jet power and assess if the jets have a negative or positive impact on the star formation in their host galaxies \citep[e.g.,][]{Mukherjee18}.


Our results support a scenario where compact radio jets are a crucial feedback mechanism during a quasar phase. Further work is now needed to decouple the relative roles of jets and winds in contributing to the total radio emission in a larger sample of quasars and their relative importance for feedback on their host galaxies. To this end, we are already working on an expanded sample selected from \citet{Mullaney13}, removing the pre-selection on sources with known outflows (see Fig.~\ref{targets}). Future, higher spatial resolution radio images (e.g. VLBI) will also help to disentangle the two forms of emission. 

\section{Conclusions}
\label{conclusions}

We have presented 0.25--1\,arcsecond resolution, 1--7\,GHz radio images and integral field spectroscopy of a sample of ten $z$$<$$0.2$ type~2 quasars ($\log [L_{\text{AGN}}$/erg\,s$^{-1}]$$\gtrsim$$45$) selected to have ionized gas outflows based on their broad [O~{\sc iii}] line widths (Fig.~\ref{targets}). Our previous work revealed that the outflows in these sources were located on $\gtrsim$kpc scales \citep{Harrison14,Harrison15}. The targets have moderate radio luminosities ($\log [L_{\text{1.4\,GHz}}$/W\,Hz$^{-1}]$=23.3--24.4) and are classified as `radio-quiet' and not as `radio AGN' using many traditional criteria \citep[e.g.,][]{Xu+99,Best&Heckman}. However, based on our \emph{UV}--\emph{FIR} SED fitting, all but one of these targets are classified `radio excess', with $\gtrsim$90\% of the total 1.4\,GHz radio luminosity not accounted for by star formation (Fig.~\ref{radio_loudness}). In this work, we have explored the origin of this radio emission and its relationship to the ionized gas distribution and kinematics. Our main conclusions are the following:

\begin{itemize}

\item Of the nine radio excess sources, we identify radio features associated with the AGN (lobes, jets, hot spots) separated by 1--25\,kpc in 7 or 8 (see Section~\ref{radio_analysis}; Fig.~\ref{radio_collage_dat}).
\item Based on the radio size--luminosity relationship, these quasars are consistent with radio-identified AGN hosting radio jets: low power compact radio galaxies or small FRI galaxies. Furthermore, the collimated appearance of many of the radio structures we observe leads us to favour radio jets as the dominant cause of the extended radio structures in the majority of our sample (see Section~\ref{sec:origin}; Fig.~\ref{radio_collage_dat}; Fig.~\ref{powervsize}).
\item For eight of the targets, we identify compact nuclear radio components ($\lesssim$400\,pc) that also appear to be dominated by processes associated with the AGN. Most of these nuclear regions have a steep 1--7\,GHz radio spectral index (i.e., $\alpha<-0.6$) which could be attributed to small-scale jets/lobes or shocked interstellar medium from quasar winds. Three have flatter spectral indices, possibly revealing a contribution from an AGN core (see Section~\ref{sec:origin}; Fig.~\ref{alpha_vs_dist}).
\item We show that there are strong indications of interactions between the observed radio jet structures and the warm ionized gas (as traced by the [O~{\sc iii}] emission-line). In particular, the two phases are spatially coincident and the radio jets/lobes we observed are co-spatial with distinct kinematic components. These observations are consistent with jet--ISM interactions resulting in galactic outflows and deflected jets (see Section~\ref{OIII_conection}; Fig.~\ref{collage_dat}; Fig.~\ref{PA}; Fig~\ref{kinematics}). 
\end{itemize}

In this work we provide evidence that compact radio jets ($\approx$1--25\,kpc) are a common feature in radiatively dominated (`radio-quiet') quasar systems and an important mechanism for driving outflows. We have demonstrated the importance of deep high-resolution radio imaging to identify the origin of the radio emission in such systems and to search for jet--ISM interactions. Our observations are in qualitative agreement with models where radio jets become stagnated as they plough into the host galaxy material, and simultaneously increase turbulence and drive large-scale outflows. Future work, in particular focusing on the energetics of both the jets and the outflows is needed to quantitatively test these models and to establish the impact of compact jets on the evolution of massive galaxies.



\section*{Acknowledgements}

We thank Tao An for graciously sharing the compiled radio data used to make Fig.~\ref{powervsize}. We thank the referee for their prompt and constructive comments. The National Radio Astronomy Observatory is a facility of the National Science Foundation operated under cooperative agreement by Associated Universities, Inc. e-MERLIN (enhanced Multi-Element Radio Linked Interferometer Network) is a National Facility operated by the University of Manchester at Jodrell Bank Observatory on behalf of the Science and Technology Facilities Council. Based on observations collected at the European Southern Observatory under ESO programmes 092.B-0062. Based on observations made with the NASA/ESA Hubble Space Telescope, obtained from the data archive at the Space Telescope Science Institute. STScI is operated by the Association of Universities for Research in Astronomy, Inc. under NASA contract NAS 5-26555. MEJ and CC acknowledge support from the IMPRS on Astrophysics at the LMU (Munich).






\bibliographystyle{mnras} %
\bibliography{jets_paper.bib} 





\bsp	
\label{lastpage}
\end{document}